\documentclass[12pt,preprint]{emulateapj}

\newcommand{\pasa}{Publ. Astron. Soc. Aust.}

\usepackage{rotating}

\shorttitle{Nuclear X-ray properties of the peculiar radio-loud hidden AGN 4C+29.30}
\shortauthors{Sobolewska et al.}

\begin{document}

\title{Nuclear X-ray properties of the peculiar radio-loud hidden AGN 4C+29.30}

\author{M. A. Sobolewska$^1$, Aneta Siemiginowska$^1$, G. Migliori$^1$,
  \L. Stawarz$^{2,\,3}$, M. Jamrozy$^3$, D. Evans$^1$, C. C. Cheung$^4$}
\affil{$^1$Harvard-Smithsonian Center for Astrophysics, 60 Garden Street,
  Cambridge, MA 02138, USA\\$^2$ Institute of Space and Astronautical Science,
  JAXA, 3-1-1 Yoshinodai, Sagamihara, Kanagawa, 229-8510, Japan\\$^3$Astronomical
  Observatory, Jagiellonian University, ul. Orla 171, 30-244 Krak\'ow, Poland
  \\$^4$ National Research Council Research Associate, National
  Academy of Science, Washington, DC 20001, resident at Naval research Laboratory,
  Washington, DC 20375, USA
}
\email{msobolewska@cfa.harvard.edu}

\begin{abstract}
We present results from a study of a nuclear emission of a nearby
radio galaxy, 4C+29.30, over a broad 0.5--200\,keV X-ray band. This study
used new {\it XMM-Newton} ($\sim 17$\,ksec) and {\it Chandra} ($\sim 300$\,ksec)
data, and archival {\it Swift}/BAT data from the 58-month catalog. The hard
($>2$\,keV) X-ray spectrum of 4C+29.30 can be decomposed into an intrinsic
hard power-law ($\Gamma\sim1.56$) modified by a cold absorber with an intrinsic
column density $N_{\rm H,\,z}\sim5\times10^{23}$\,cm$^{-2}$, and its reflection
($|\Omega/2\pi| \sim 0.3$) from a neutral matter including a narrow iron
K$\alpha$ emission line at the rest frame energy $\sim 6.4$\,keV. The reflected
component is less absorbed than the intrinsic one with an upper limit on the
absorbing column of $N^{\rm refl}_{\rm H,\,z}<2.5\times 10^{22}$\,cm$^{-2}$.
The X-ray spectrum varied between the {\it XMM-Newton} and {\it Chandra}
observations. We show that a scenario invoking variations of the normalization
of the power-law is favored over a model with variable intrinsic column density.
X-rays in the 0.5--2\,keV band are dominated by diffuse emission modeled with
a thermal bremsstrahlung component with temperature $\sim 0.7$\,keV, and contain
only a marginal contribution from the scattered power-law component. We
hypothesize that 4C+29.30 belongs to a class of `hidden' AGN containing a
geometrically thick torus. However, unlike the majority of them, 4C+29.30 is
radio-loud. Correlations between the scattering fraction and Eddington luminosity
ratio, and the one between black hole mass and stellar velocity dispersion, imply
that 4C+29.30 hosts a black hole with $\sim 10^8$\,M$_{\odot}$ mass.
\end{abstract}


\section{Introduction}

4C+29.30 is a low-redshift ($z=0.0647$) radio source with a moderate radio luminosity
$\sim 10^{42}$\,erg\,s$^{-1}$ (Siemiginowska et al. 2012) hosted by an elliptical galaxy.
It was first studied in the radio and optical bands by van Breugel et al. (1986).
Observed complex radio morphology was resolved into jets, knots, lobes, and a diffuse tail.
A number of optical emission lines were measured, and their relative
intensities indicated photoionization by a non-stellar continuum attributed to the
central active galactic nucleus (AGN). The radio emission and the optical line emitting
gas were found to be strongly linked suggesting interaction between the radio source and
its environment. In addition, shells and dust lanes observed in the optical band (van Breugel
et al. 1986) provide evidence of a past merger with a gas-rich disk galaxy which possibly
triggered the rejuvenated AGN activity. Subsequently, follow-up radio studies revealed
radio structures with diverse spectral ages suggesting an intermittent nature of the radio
source in 4C+29.30. Jamrozy et al. (2007) presented evidence for a large scale extended
relic radio emission most probably due to an earlier cycle of activity of the source
$\gtrsim 200$\,Myr ago. The age of a small-scale radio structure embedded in the extended relic
radio emission was estimated at $\lesssim 100$\,Myr, with the inner double knots of a spectral
age of $\lesssim 33$\,Myr (Jamrozy et al. 2007). Liuzzo et al. (2009) resolved the very central
region of the source into two nuclear knots with spectral ages of $\sim 15$\,yr and $\sim 70$\,yr.
4C+29.30 was cataloged as an
infrared point-source by the IRAS, WISE and 2MASS surveys (NASA/IPAC Infrared Science Archive).

The first detection of the source in the X-ray band occurred during a snap-shot 8\,ksec {\it Chandra}
observation (Gambill et al. 2003). The short exposure time and few detected counts prevented detailed
spectral analysis. Nevertheless, this pilot observation hinted towards a complex X-ray morphology
composed of a nucleus, hotspots, a putative jet, and diffuse emission. The source was not a target
of X-ray pointings prior to the first {\it Chandra} observation, without detection in, e.g., the ROSAT
All Sky Survey, probably due to significant intrinsic absorption (Gambill et al. 2003,
Siemiginowska et al. 2012). In 2008, we observed 4C+29.30 with {\it XMM-Newton} for $\sim 45$\,ksec.
This observation was followed by our deep 300\,ksec {\it Chandra} exposure (Siemiginowska et al. 2012,
hereafter Paper I) which clearly revealed diverse X-ray emitting structures, many of which
correspond to those in radio or optical bands, and others are intrinsic to the X-ray band.
The soft 0.5--2\,keV X-ray image showed northern and southern lobes, hotspots, southern jet, and
thermal diffuse emission in the center. Emission in the hard X-ray band (above 2\,keV) was instead
dominated by the active nucleus of 4C+29.30. In Paper I, our discussion focused on the spectral properties
of the soft X-ray components, the structure of the outflow, and interactions of the X-ray jet with
the interstellar medium. We emphasized similarities between 4C+29.30 and famous radio galaxies showing
jet-ambient medium interactions, i.e. M87 (e.g., Million et al. 2010), NGC 1275
(e.g., Fabian et al. 2011), and Cen A (e.g., Morganti et al. 2010,
and references therein).

In the present paper, we focus on detailed modeling of the X-ray core emission to uncover
the nature of the AGN powering 4C+29.30. We analyze new {\it Chandra} data and archival
{\it Swift}/BAT data from the 58-months catalog (Baumgartner et al. 2010), and thus we study
the core emission over a broad 0.5--200\,keV X-ray band. We apply our best fit model to
our new {\it XMM-Newton} observations of 4C+29.30 taken $\sim 2$ years before our {\it Chandra}
observation. We search for signatures of the X-ray spectral variability between the {\it Chandra}
and {\it XMM-Newton} observations. The paper is organized as follows. In Section 2 we describe
our data and provide details on the preparation of the spectra for modeling. In Section 3 we test
various spectral scenarios for AGN X-ray activity. We compare the best fitting {\it Chandra} and
{\it XMM-Newton} models in order to study the origins of possible spectral variability. In Section
4 we present the discussion of our results indicating that 4C+29.30 is a strongly absorbed
intrinsically X-ray variable AGN with Seyfert 2 type of activity and a peculiar obscuring geometry,
probably due to a geometrically thick torus similar to that reported for several {\it Swift}/BAT
selected AGN (e.g., Ueda et al. 2007; Winter et al. 2009a,b). Finally, in Section 5 we summarize
and conclude our study. Throughout the paper we adopt $\Omega_\Lambda=0.73$, $\Omega_{\rm M}=0.27$
and $H_{\rm 0}=70$\,km\,s$^{-1}$\,Mpc$^{-1}$ for the flux-luminosity transformations.

\section{X-ray observations}

\subsection{\it XMM-Newton}

Our {\it XMM-Newton} observation of 4C+29.30 was performed on Apr 11,
2008.  It was split in two pointings lasting $\sim 45.7$\,ksec (ID:
0504120101) and $\sim 9.7$\,ksec (ID: 0504120201). No science data
were produced during the second pointing.  The data reduction of the
$45.7$\,ksec long observation was performed following the standard
procedure\footnote{http://xmm.esac.esa.int/sas/current/documentation/threads/}
with the {\tt XMM-SAS} v. 11.0.0 package using the updated
calibration files (Oct 2011).  We constructed light curves at energies
above 10\,keV to identify periods of high background flaring.  After
cleaning the data, we obtained the following net exposures and 0.5--10\,keV
observed count rates: $\sim 17$\,ksec and
$(6.3 \pm 0.2)\times10^{-2}$\,cnt\,s$^{-1}$ for
the EPIC/PN camera, $\sim 23$\,ksec and $(1.75\pm 0.09)\times10^{-2}$
cnt s$^{-1}$ for MOS1, and $\sim 23$\,ksec and $(2.03 \pm 0.09)\times10^{-2}$
cnt s$^{-1}$ for MOS2. In this paper we will limit the discussion
to the EPIC/PN data which provide better statistic than the MOS data.
However, the data taken by the EPIC/MOS1 and MOS2 have been checked
for consistency.

\begin{figure*}
\begin{center}
\includegraphics[height=5.2cm]{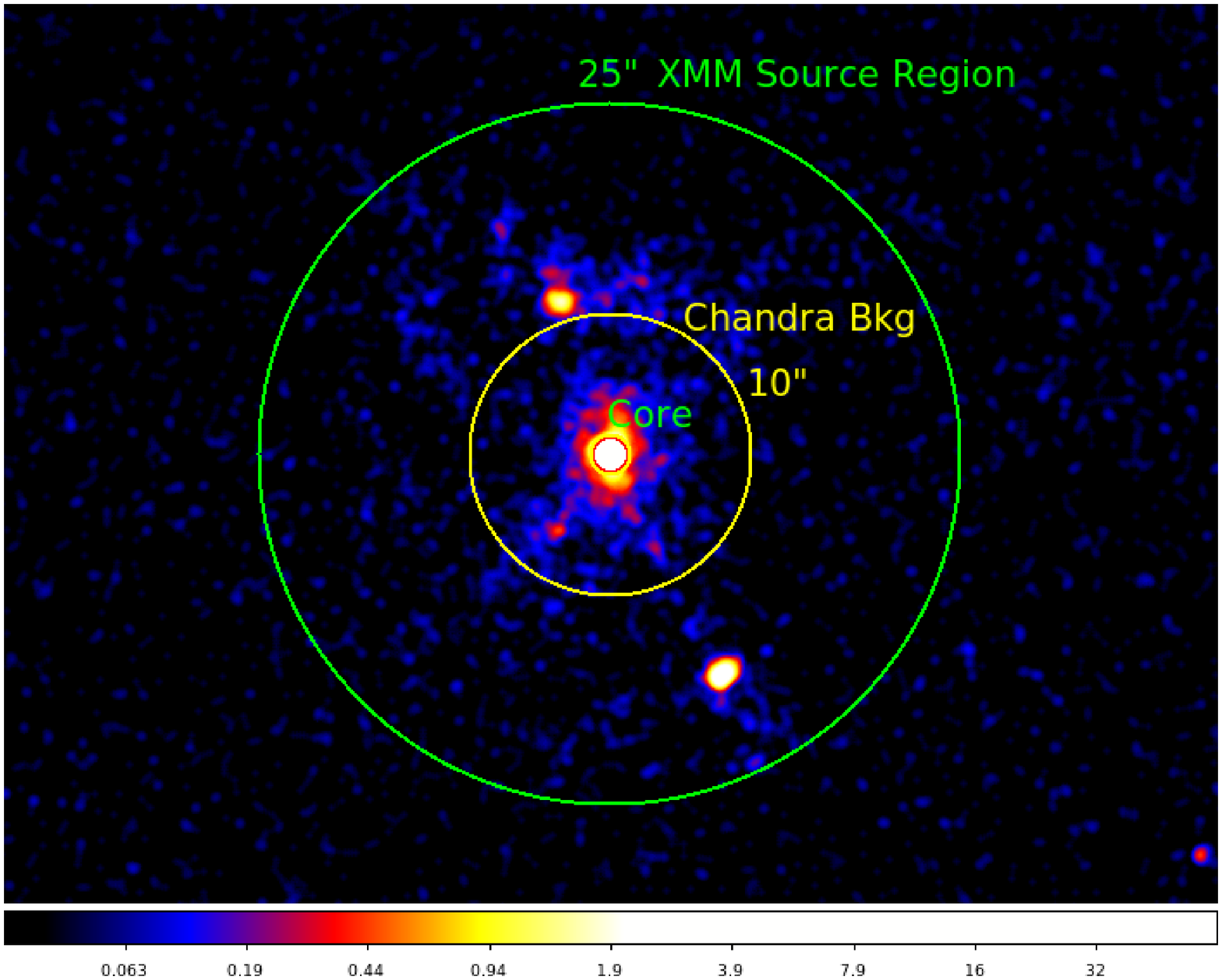}~
\includegraphics[height=5.2cm, bb=55 7 517 433,clip]{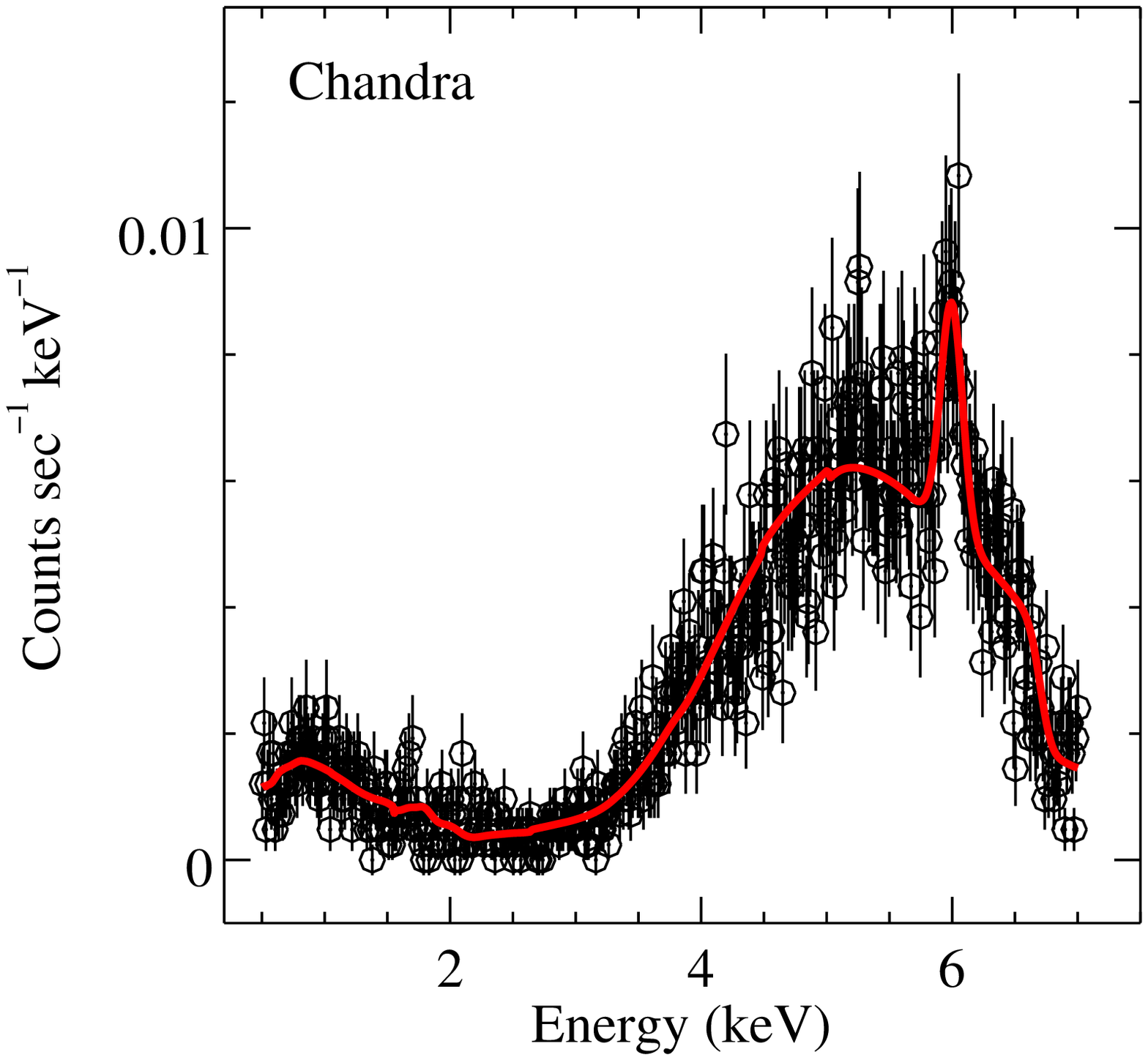}
\includegraphics[height=5.2cm, bb=90 7 517 433,clip]{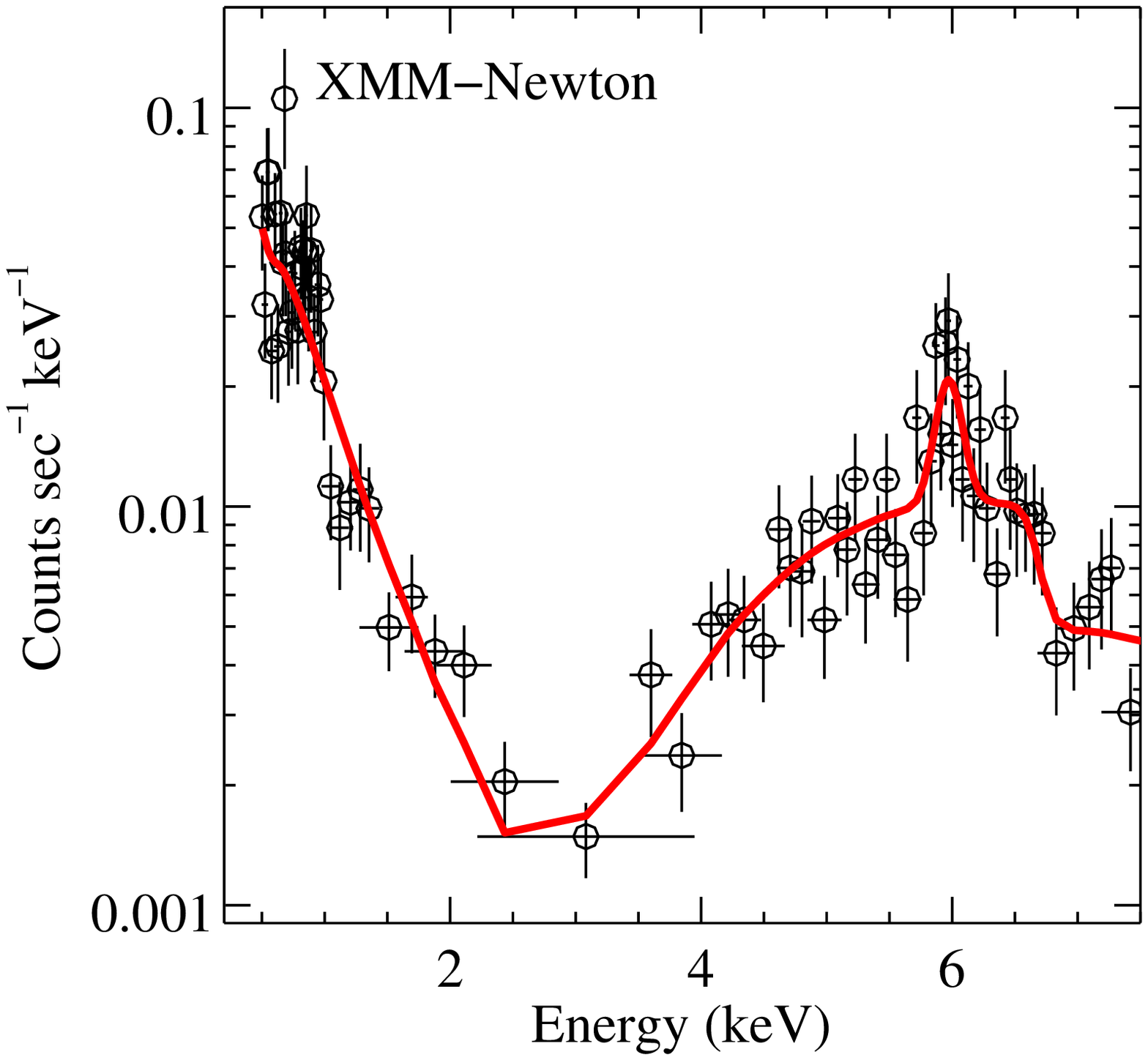}
\caption{Left: Extraction regions for 4C+29.30 core spectra in the {\it Chandra} and
  {\it XMM-Newton} observations (1.25$\arcsec$ magenta and 25$\arcsec$ green circles,
  respectively), and 10$\arcsec$ {\it Chandra} background overplotted on top of
  our deep {\it Chandra} image. The ACIS-S {\it Chandra} image has
  been filtered and includes only 0.5-7\,keV photons. The original ACIS pixels
  were binned to 0.123$\arcsec$ and smoothed with Gaussian function of $\sigma=0.615\arcsec$.
  Middle: Unbinned {\it Chandra} spectrum of the 1.25$\arcsec$core region and folded
  Model 3 (see Sec.~3.1). Right: {\it XMM-Newton} spectrum of the 25$\arcsec$ source
  region and folded Model 3b (see Sec.~3.2).}
\label{fig:regs}
\end{center}
\end{figure*}

The source 0.5--10\,keV PN spectrum was extracted from a circular region
(centered at R.A. 08:40:02.276, Decl. +29:49:02.47;
J2000.0) of 25$\arcsec$ radius (Figure~\ref{fig:regs}), and the background
from a circular region located on the same CCD (50$\arcsec$ radius; not
indicated in Figure~\ref{fig:regs}).  The response matrices were created
using the SAS commands \texttt{rmfgen} and \texttt{arfgen}. The source data
are not piled-up.  We used \texttt{specgroup} to set spectral groups such
that each group/bin contains a minimum signal-to-noise ratio of 3.

\subsection{\it Chandra}

Deep {\it Chandra} ACIS-S imaging observations of 4C+29.30 were
performed in Feb 2010, i.e. $\sim 2$ years after our {\it XMM-Newton}
observation. The approved 300\,ksec observation was split into four
separate pointings (OBSIDs 11688, 11689, 12106, 12119) that sum to
a total exposure of 286.4\,ksec (see
Paper I for details of the observations and data
reduction). In this paper we use the four combined\footnote{Because
the data were collected on the same part of the ACIS-S detector and
spanned a relatively short time interval, we used CIAO 4.3 script {\tt
combine\_spectra} to combine individual spectra and response files
into one spectrum for this analysis.} 0.5--7\,keV X-ray spectra
($5328\pm73$ total counts) of the core region (Figure~\ref{fig:regs})
defined as a circle with radius of 1.25$\arcsec$ centered at the position
of the nucleus determined based on the hard band {\it Chandra} X-ray image
(R.A. 08:40:02.345, Decl. +29:49:02.61, J2000.0; Paper I), together with
the appropriate calibration response files. The background ($25\pm5$ counts)
was extracted from an annulus centered on the core position with an inner
radius of 1.5$\arcsec$ and an outer radius of 10$\arcsec$. We use CIAO
version 4.3 tools for the data analysis.

\subsection{{\it Swift}/BAT}

We extend the {\it Chandra} and {\it XMM-Newton} energy bandpasses by
considering the publicly available 14--195\,keV (8-channel) hard X-ray
{\it Swift}/BAT spectrum from the 58-months catalog (between Nov 2004
and Sep 2008; Baumgartner et al. 2010) together with
the corresponding response matrix {\tt diagonal\_8.rsp}. We do not
detect any long term variability signal in the {\it Swift}/BAT lightcurve (normalized
excess variance $\sigma^2_{\rm nxs} < 0$; e.g., Vaughan et al. 2003).
Either way, integrating the data over the period of $\sim 5$ years
would smear any hard X-ray spectral variations. Thus, the {\it Swift}/BAT data are
representative of the average hard X-ray emission from 4C+29.30.

\section{Modeling}

In this section we use our deep {\it Chandra} observation to establish
a spectral model for the 4C+29.30 core continuum. We use the {\it
 Swift}/BAT data to check the consistency of the model with X-ray
data outside the {\it Chandra} energy bandpass. Finally, we apply
the best fitting {\it Chandra} model to the {\it XMM-Newton}
observation taken $\sim 2$ years prior to the {\it Chandra} pointing,
and we discuss the X-ray spectral variability of the source. We
perform the fits in {\it Sherpa} (Freeman et al. 2001) using CSTAT statistic.

The {\it Swift}/BAT data were averaged over 58 months of observations, which
potentially might introduce a discrepancy between the {\it Chandra} (or {\it
  XMM-Newton}) and {\it Swift}/BAT data due to the X-ray flux variability.
{\it Swift}/BAT composite spectra contain fractional counts with Gaussian statistic
and are background subtracted, which prevents the CSTAT fitting.
\footnote{http://heasarc.nasa.gov/docs/swift/analysis/threads/batspectrumthread.html}
For this reason, we find best fit of each model to the {\it Chandra} (or {\it XMM-Newton})
data only. Then we apply the best fit model to the {\it Swift}/BAT data using $\chi^2$
statistic with variance supplied with the data
{\tt chi2xspecvar}, allowing for only one free parameter, the normalization constant,
$C_{\rm BAT}$.

Table~\ref{tab.mname} contains the description of all models used in
the paper in the {\tt XSPEC} (Arnaud 1996) and {\tt Sherpa} terminology. In Table~\ref{tab.mpar} we list
the best fit model parameters and the resulting fit statistics. All models
considered in this section include Galactic absorption ({\tt
  phabs}) with a column density fixed at $N_{\rm H} = 3.98\times
10^{20}$\,cm$^{-2}$ (Dickey \& Lockman 1990).  Errors on the spectral
parameters correspond to the 68\% confidence level (1$\sigma$) for
one significant parameter.

A region of the diffuse emission surrounding the nucleus can be clearly
identified in the soft 0.5--2\,keV {\it Chandra} image (`Center' in Figure~2 in
Paper I). In Paper I, we showed that a
thermal bremsstrahlung model provided a better fit to the
spectra extracted from this region than a power-law model. Thus,
all our models include a thermal bremsstrahlung component with the
temperature and normalization allowed to vary.

We modeled the {\it Chandra} and {\it XMM-Newton} background spectra
by fitting appropriate background regions (Figure~\ref{fig:regs}) with a
model consisting of a Galactic absorption, thermal bremsstrahlung and
power-law. The background model parameters are listed in
Table~\ref{tab.mpar}. The parameters differ because the {\it XMM-Newton}
background was collected from a region at the distance $> 25\arcsec$
from the nucleus.

\subsection{{\it Chandra}}

We considered several models for the nuclear X-ray emission of
4C+29.30 observed with {\it Chandra}.

\begin{figure}
\begin{center}
\includegraphics[height=5.2cm]{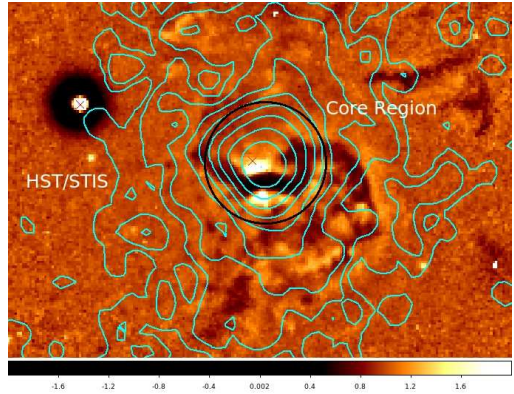}
\caption{An unsharp mask optical HST image of 4C+29.30 with overplotted {\it Chandra}
  X-ray contours (cyan; Paper I). The crosses indicate the optical SDSS identified counterparts.
  The 1.25$\arcsec$ {\it Chandra} core extraction region is indicated (black circle).
  A dust lane across the optical nucleus is clearly visible.}
\label{fig:opt}
\end{center}
\end{figure}

\begin{figure*}
\vspace{1.2cm}
\begin{center}
\includegraphics[bb=80 -50 1055 282,scale=0.25]{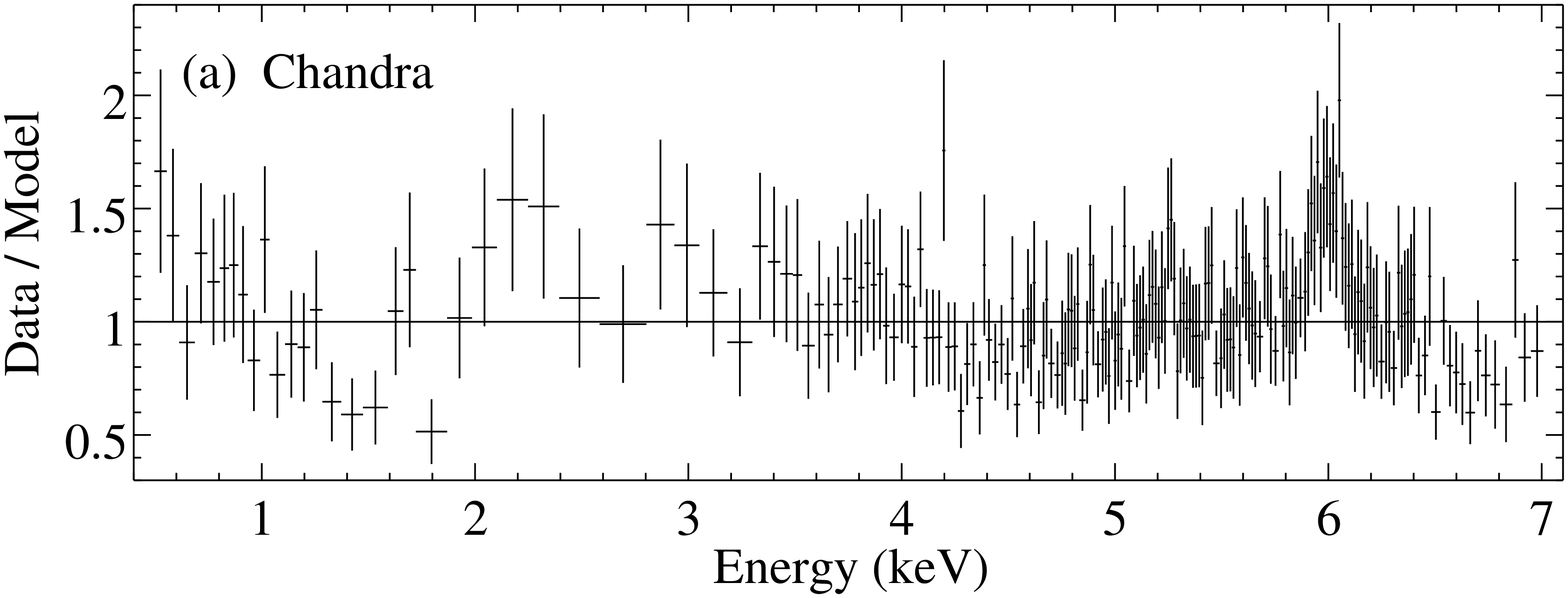}
\includegraphics[bb=80 -50 1055 282,scale=0.25]{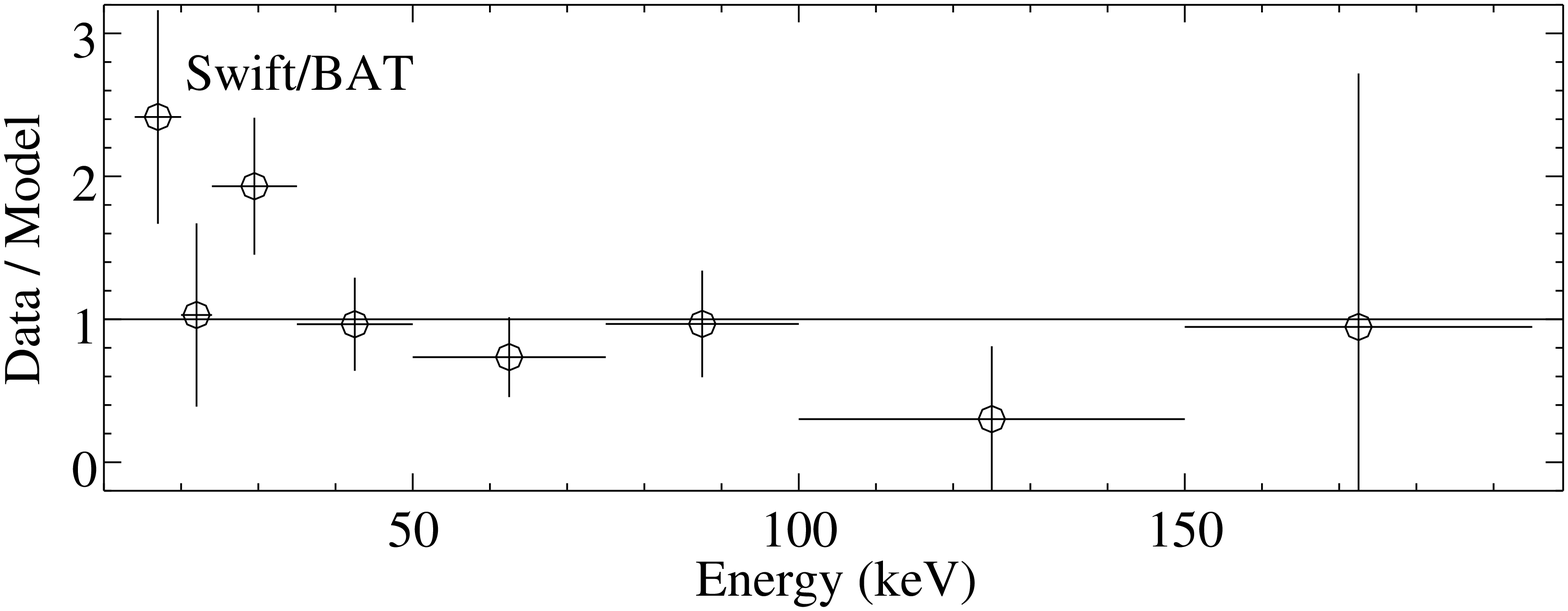}\\
\includegraphics[bb=80 -50 1055 282,scale=0.25]{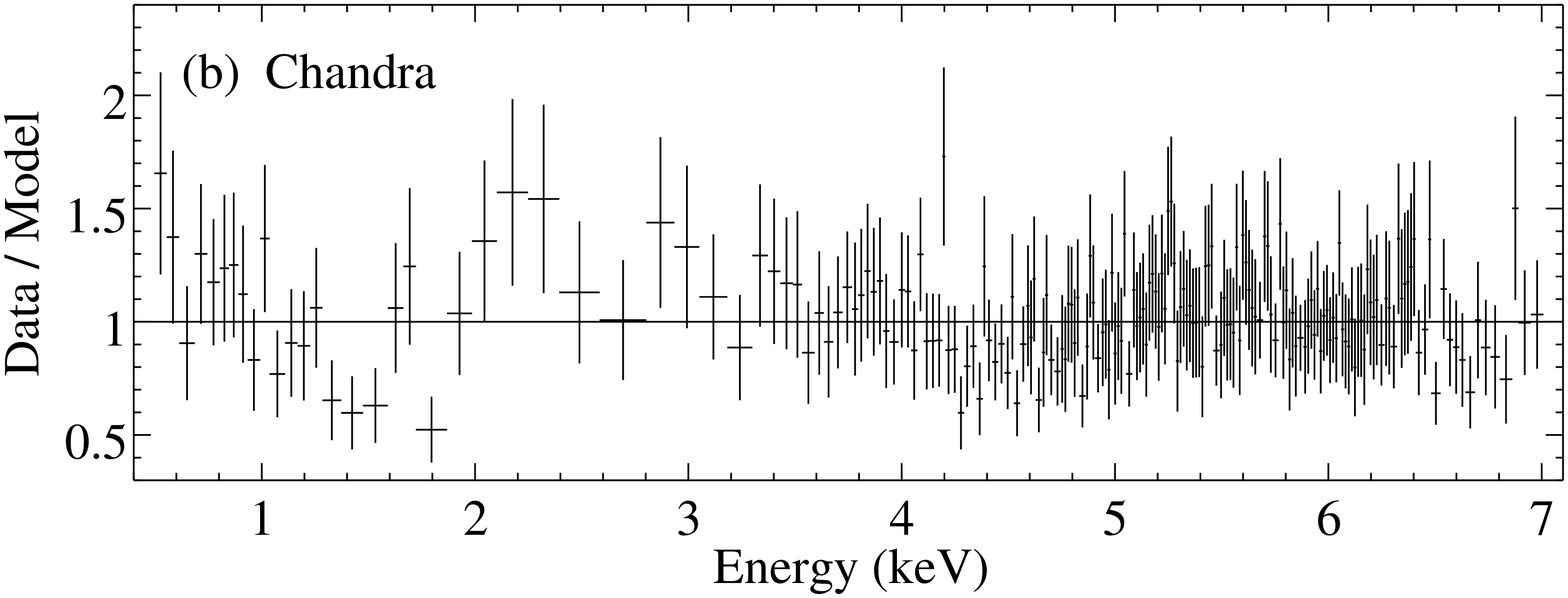}
\includegraphics[bb=80 -50 1055 282,scale=0.25]{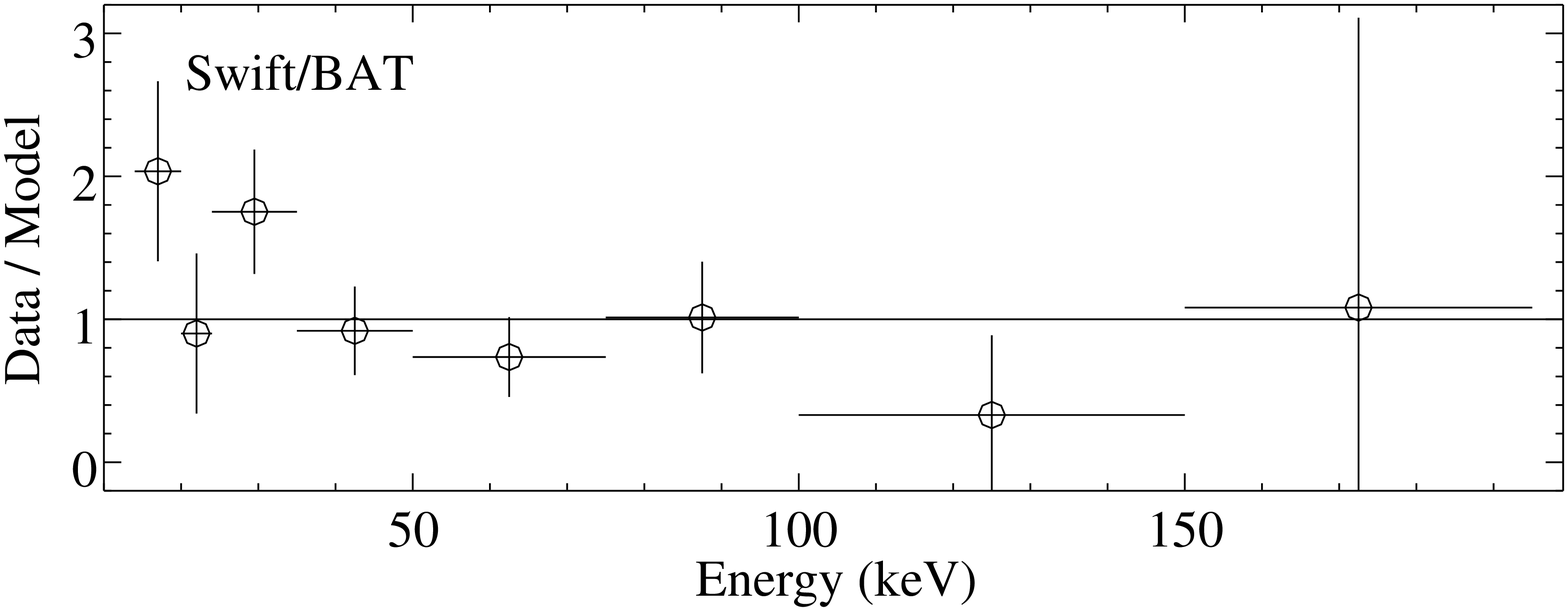}\\
\includegraphics[bb=80 -50 1055 282,scale=0.25]{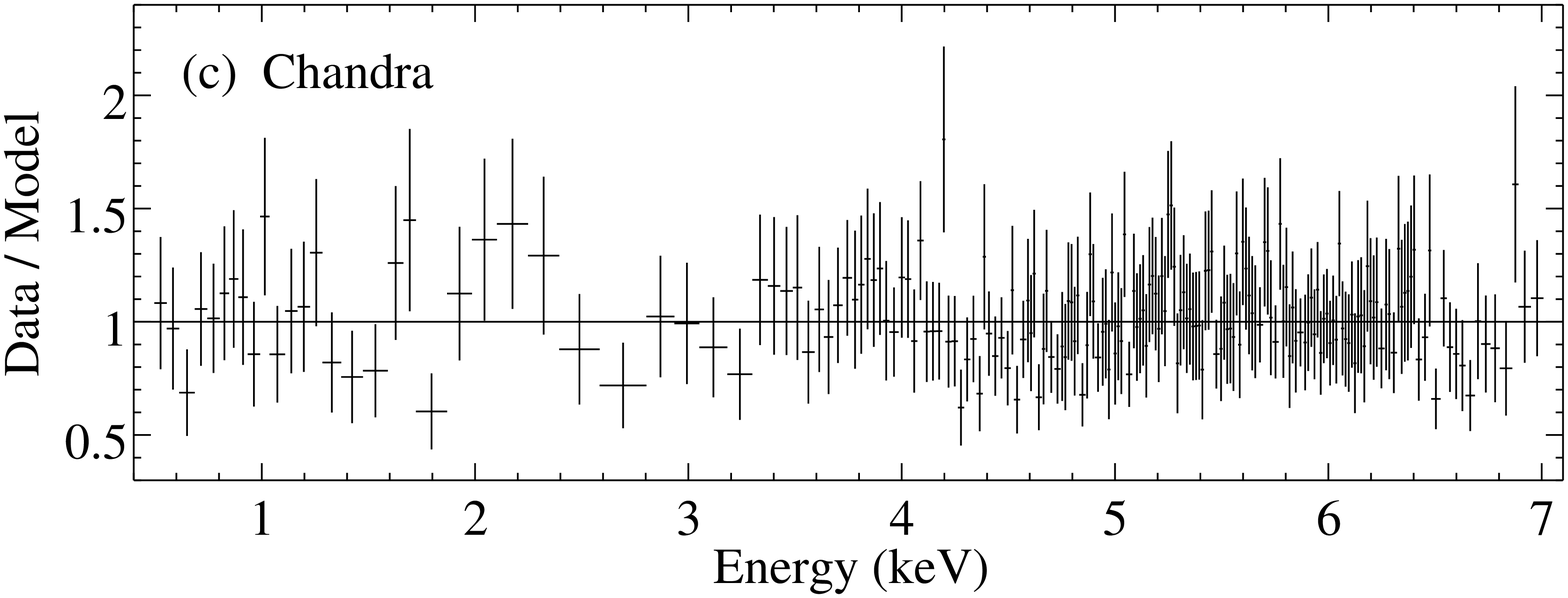}
\includegraphics[bb=80 -50 1055 282,scale=0.25]{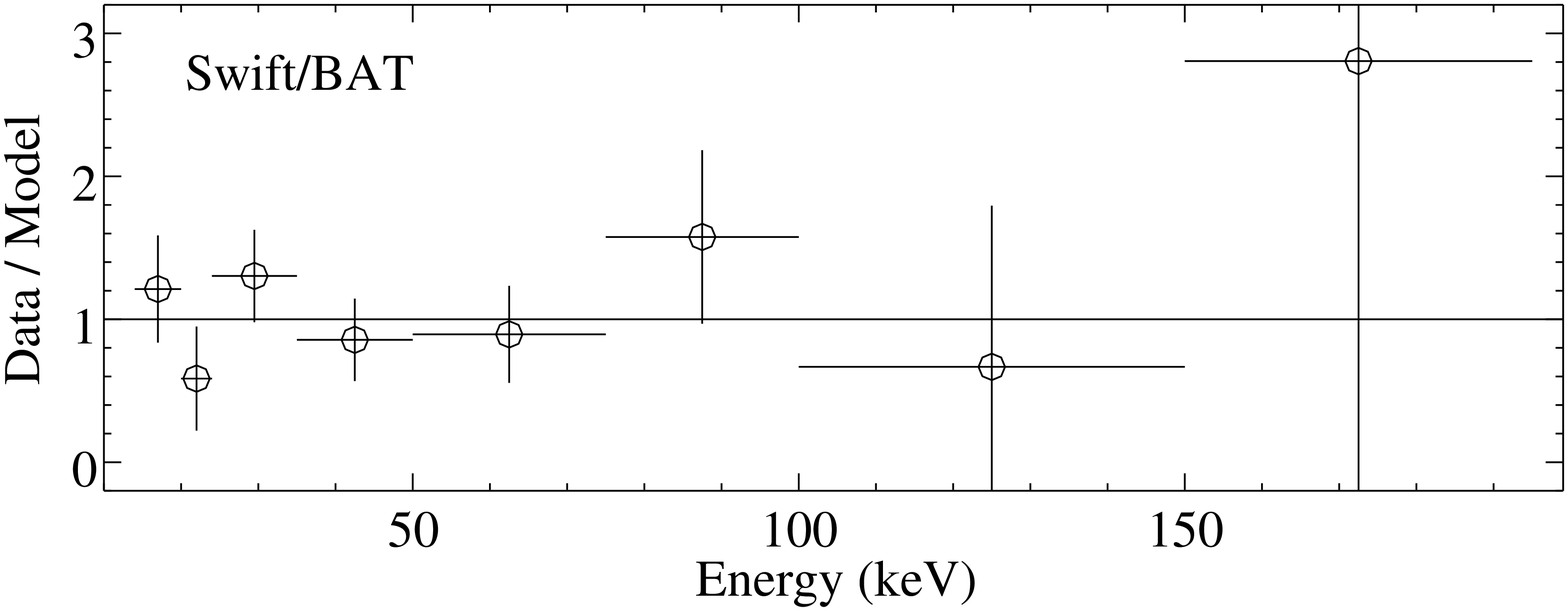}
\caption{Ratio of the {\it Chandra} and {\it Swift}/BAT data to different spectral models.
{\it Chandra} data were binned for the clarity of the plot.
(a) Model 1 -- in the {\it Chandra} band clear signatures of iron K$\alpha$
line are visible around 6\,keV ($\sim 6.4$\,keV in the rest frame); resulting
photon index is in disagreement with the {\it Swift}/BAT data.
(b) Model 2 -- the $\sim 6$\,keV residuals disappear after including a Gaussian
line in the model; the photon index is still too low to agree with
the {\it Swift}/BAT data.
(c) Model 3 with a Gaussian line and reflection fits well both
the {\it Chandra} and {\it Swift}/BAT data.}
\label{fig:ratcb}
\end{center}
\end{figure*}

{\it Model 1.} First, we tried a simple parametrization using a cut-off power-law
function ({\tt zhighect*zpowerlw} in {\tt XSPEC} terminology; in the following
we will refer to this component as CPL or `a power-law'). The {\tt zhighect}
model is defined as $\exp[E_{\rm 1}/E_{\rm fold} - E\times(1+z)/E_{\rm fold}]$. We assumed
$E_1 = 0$. The E-folding energy, $E_{\rm fold}$ could not be constrained by our
data and we fixed it at 300\,keV.
We added an intrinsic absorber with a variable column density
({\tt zphabs}), motivated by the presence of a dust line crossing the nuclear region,
as revealed by the HST image (Figure~\ref{fig:opt}). We were not able to obtain a good
fit with this model. The plot of the model-to-data ratio revealed
residuals in the 5--7\,keV band suggesting the presence of an iron K$\alpha$
emission line (Figure~\ref{fig:ratcb}a). These residuals forced the photon index of the
intrinsically absorbed ($N_{\rm H,\,z} \sim 4 \times 10^{23}$\,cm$^{-2}$) continuum
to converge to an unusually low value, $\Gamma = 0.84^{+0.11}_{-0.27}$.

{\it Model 2.} In Model 2 we added a Gaussian line ({\tt zgauss}) which reduced the
5--7\,keV residuals significantly (Figure~\ref{fig:ratcb}b). The resulting
line was narrow, $\sigma=(0.07\pm0.03)$\,keV, and centered at the rest frame energy
of $E_{K\alpha} = (6.36\pm0.02)$\,keV. These parameters suggest that the line originates
in a neutral material far from the black hole, e.g., at the outskirts of an accretion disk,
or a dusty torus. The earlier short 8\,ksec {\it Chandra} observation contained
too few counts to detect the line or put an upper limit on its EW (Gambill et al. 2003).
Thus, this is the first detection of the iron K$\alpha$ line in 4C+29.30.
The observed equivalent width of the line (i.e. relative to the intrinsically absorbed
continuum) was EW$_{\rm o}=192^{+54}_{-18}$\,eV, while the intrinsic one (i.e. relative
to the unabsorbed continuum) was EW$_{\rm i}=91^{+27}_{-18}$\,eV. The photon index in
Model 2 was still unusually low, $\Gamma=0.97^{+0.07}_{-0.03}$, and the intrinsic
absorption remained strong with $N_{\rm H,\,z} \sim 5 \times 10^{23}$\,cm$^{-2}$. The
temperature of the diffuse emission, $kT = 5.2^{+3.0}_{-1.5}$\,keV, was in agreement with that
measured in Model 1 ($kT = 6.4^{+6.5}_{-2.1}$\,keV), but considerably higher than
the temperature of the background thermal bremsstrahlung component, $kT = (0.63\pm0.04$)\,keV.

{\it Model 3.} The detection of the iron K$\alpha$ line suggested that the total spectrum may
contain a certain contribution of the associated reflected emission. Hence, we considered
the reflection model with a slab geometry, {\tt pexriv} of Magdziarz \& Zdziarski (1995),
i.e. a reflection from an accretion disk. We first assumed that the reflected component is not
intrinsically absorbed, contrary to the case of the incident CPL. This corresponds to a
scenario in which the CPL continuum is observed through an obscuring material, e.g., a dusty
torus, while the reflected emission arrives to the observer unobscured. With default
settings the {\tt pexriv} model computes both the incident CPL and the reflected component.
However, in order to apply the intrinsic absorber only to the incident CPL, we leave the CPL
component in the model, and choose a setting in which the {\tt pexriv} model returns only the
reflected component. Consequently, the reflection amplitude, $\Omega/2\pi$, was fitted over
a negative range, which physically corresponds to a reflection component with amplitude
$|\Omega/2\pi|$.

We linked the photon index and the normalization of
the {\tt pexriv} model to those of CPL. The {\tt pexriv} reflection does not account for
the iron line emission and so we kept the {\tt zgauss} component explicitly in the model.
The viewing angle, $i$, is not known for our source. However, an upper limit can be calculated
based on the jet-to-counterjet flux ratio, $J$, at 5 GHz (e.g., Urry \& Padovani 1995).
Sambruna et al. (2004) and Liuzzo et al. (2009) give $J > 29$ and $J > 50$, implying
$i<60^{\circ}$ (for discrete knots). We fixed $i$ at 45$^\circ$, assuming that the jet in
4C+29.30 is well aligned with the axis of the accretion disk. We fixed the abundances of
elements at the Solar value.

The resulting data-to-model residuals are presented in Figure~\ref{fig:ratcb}c.
The ionization parameter in the {\tt pexriv} model converged to zero.
The iron line remained narrow, and centered at the rest
frame energy of $E_{{\rm K}\alpha} = (6.36\pm0.01)$\,keV. Its observed EW$_{\rm o}=173^{+62}_{-15}$\,eV
(intrinsic EW$_{\rm i}=71^{+27}_{-15}$\,eV) was lower than in Model 2.  Thus, the properties of the iron line
and the reflected emission support the hypothesis that both components originate
in a neutral matter located far away from the black hole. The 
power-law component with the photon index of $\Gamma=1.56^{+0.15}_{-0.04}$ was absorbed with
an intrinsic column density $N_{\rm H,\,z} \sim 5 \times 10^{23}$\,cm$^{-2}$.
The reflection amplitude yielded an intermediate value of
$|\Omega/2\pi| = 0.32^{+0.05}_{-0.13}$. The temperature of the soft diffuse emission, $kT =
0.71^{+0.15}_{-0.12}$\,keV, was significantly lower than that in Models 1 and 2 due to
the contribution of the reflected component to the soft X-ray band, and slightly higher than
(but consistent with) the temperature of the background thermal component.
As a final model check we replaced the thermal bremsstrahlung component with
an {\tt apec} model. We were able to establish only an upper limit on the value of metal
abundances, $A \lesssim 0.01$.
It is generally known that low abundances
might be an artifact resulting from fitting the mixed temperature
plasma where these mixed temperature components cannot be spectrally or
spatially resolved (see e.g. Kim 2012 in context of abundances in
hot ISM of elliptical galaxies). This problem with regard to the data of 4C+29.30
was considered in more details in Siemiginowska et al. (2012). Similarly to their
approach we decided to keep the thermal bremsstrahlung component in the model
as a good approximation.

{\it Model 4.} Next, we allowed for the second intrinsic absorber altering
the reflected component with a column density $N^{\rm refl}_{\rm H,\,z}$, in addition to
the one modifying the intrinsic CPL emission. This model, however, converged to Model 3 with
$N^{\rm refl}_{\rm H,\,z}=1.2^{+1.3}_{-1.2}\times 10^{22}$\,cm$^{-2}$, consistent
with zero.

{\it Model 5.} Motivated by the studies of other Seyfert 2 type galaxies
(e.g., Evans et al. 2007, 2010; Grandi et al. 2007) we tested
if a partial covering model (representative for, e.g., a patchy torus) could
provide an alternative description of our data.
We assumed that only a fraction, $f_{\rm pc}$ (allowed to vary between 0 and 1), of the
power-law emission is passing through an obscuring matter, while
the remaining fraction (1-$f_{\rm pc}$) arrives to the observer
unabsorbed. This model also converged to Model 3 with $f_{\rm pc}=1.0000_{-0.0021}$.

{\it Model 6.} Next, we checked if a certain fraction, $f_{\rm sc}$, of the 
intrinsic CPL may contribute to the soft emission below 1\,keV due to scattering. We found
that the scattered component was not formally required by the data with an upper limit on
the scattering fraction $f_{\rm sc} < 0.25 \times 10^{-2}$, thus Model 6 converged to Model 3.

In summary, we conclude that Model 3 accounts properly for the broad band X-ray continuum measured
by {\it Chandra}. In this model, the nuclear radiation described by a cut-off
power-law with a hard spectrum, $\Gamma\sim1.56$, is heavily intrinsically absorbed by a cold gas
with $N_{\rm H,\,z} \sim 5\times 10^{23}$\,cm$^{-2}$. The reflected radiation arrives to the
observer unabsorbed, or significantly less absorbed, than the intrinsic emission. The upper limit
on the absorbing column density of the reflected component is
$N^{\rm refl}_{\rm H,\,z}<2.5\times10^{22}$\,cm$^{-2}$ based on the Model 4 test. The reflecting
material is neutral and located far from the black hole, as suggested by the width and energy
of the fluorescent iron K$\alpha$ line, and the ionization parameter of the reflected component.
When this best fit model is applied to the {\it Swift}/BAT data, we find a normalization
constant between the two data sets, $C_{\rm BAT} = 0.68\pm0.10$.

The right panels in Figure~\ref{fig:ratcb} show data-to-model ratios (Models 1--3) in the
{\it Swift}/BAT energy range. The figures indicate that the Model 3 scenario is favored and
we plot the model and its components in Figure~\ref{fig.eeuf}a.

\subsection{{\it XMM-Newton} data}

In this section we apply the best fitting {\it Chandra}
model (Model 3) to the {\it XMM-Newton} data.

\begin{figure*}
\begin{center}
\vspace{1.2cm}
\includegraphics[bb=80 -115 1055 282,scale=0.25]{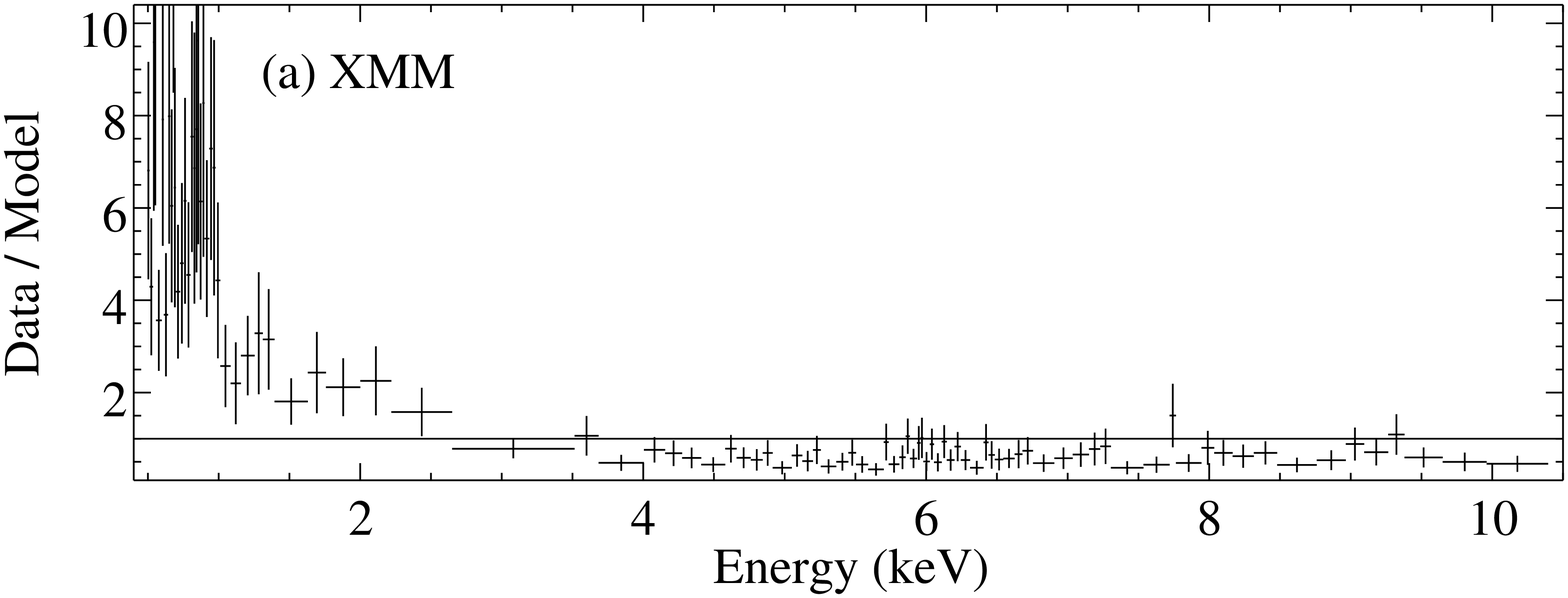}
\includegraphics[bb=80 -115 1055 282,scale=0.25]{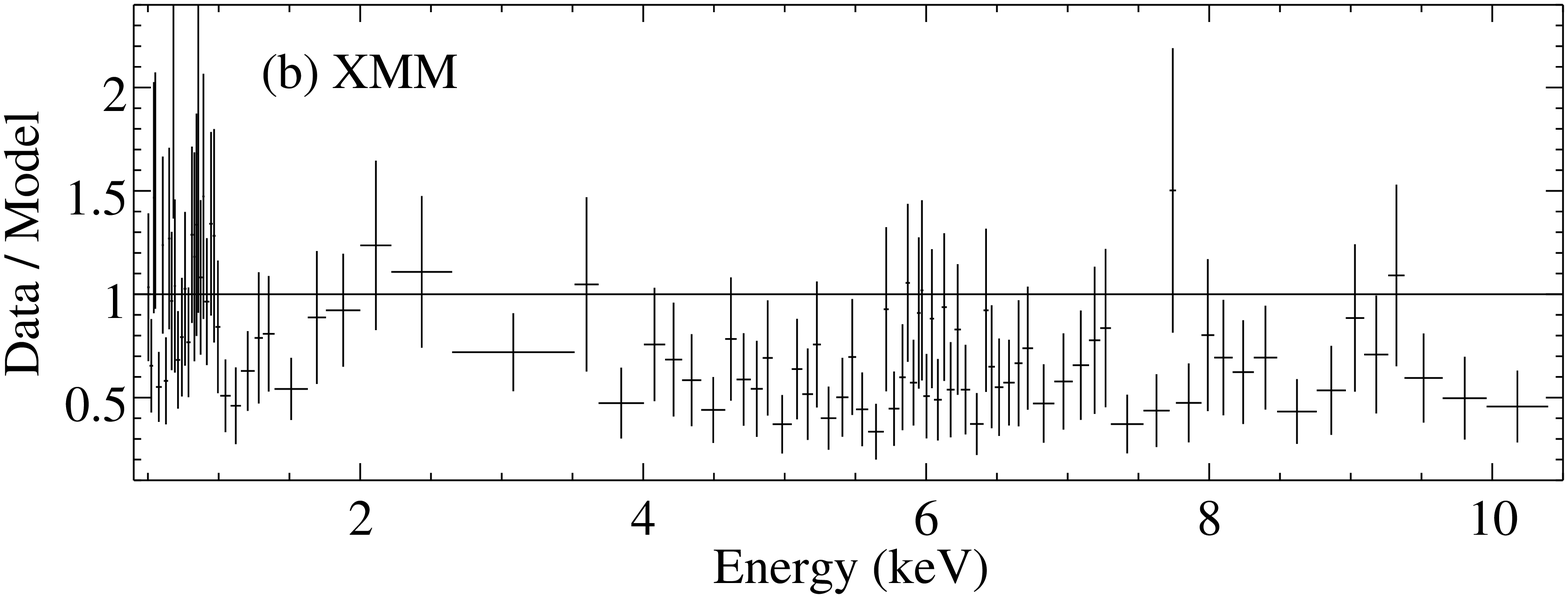}\\
\includegraphics[bb=80 -50 1055 282,scale=0.25]{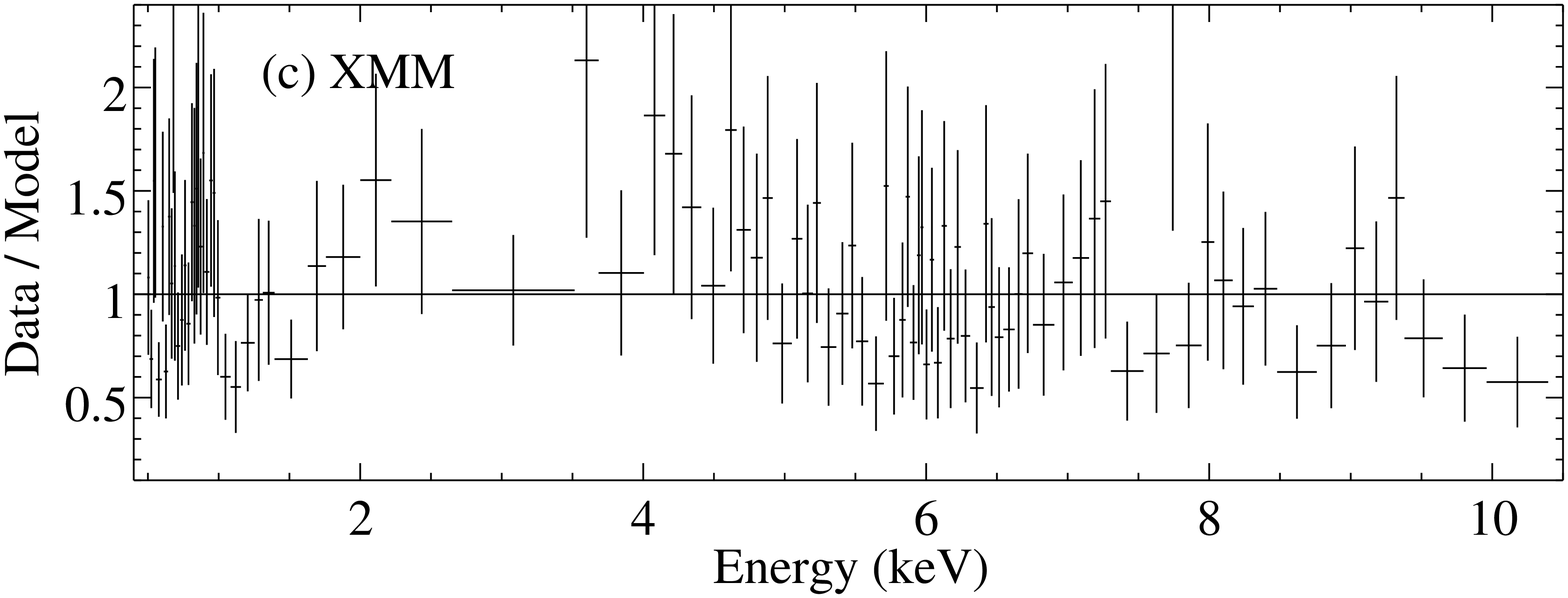}
\includegraphics[bb=80 -50 1055 282,scale=0.25]{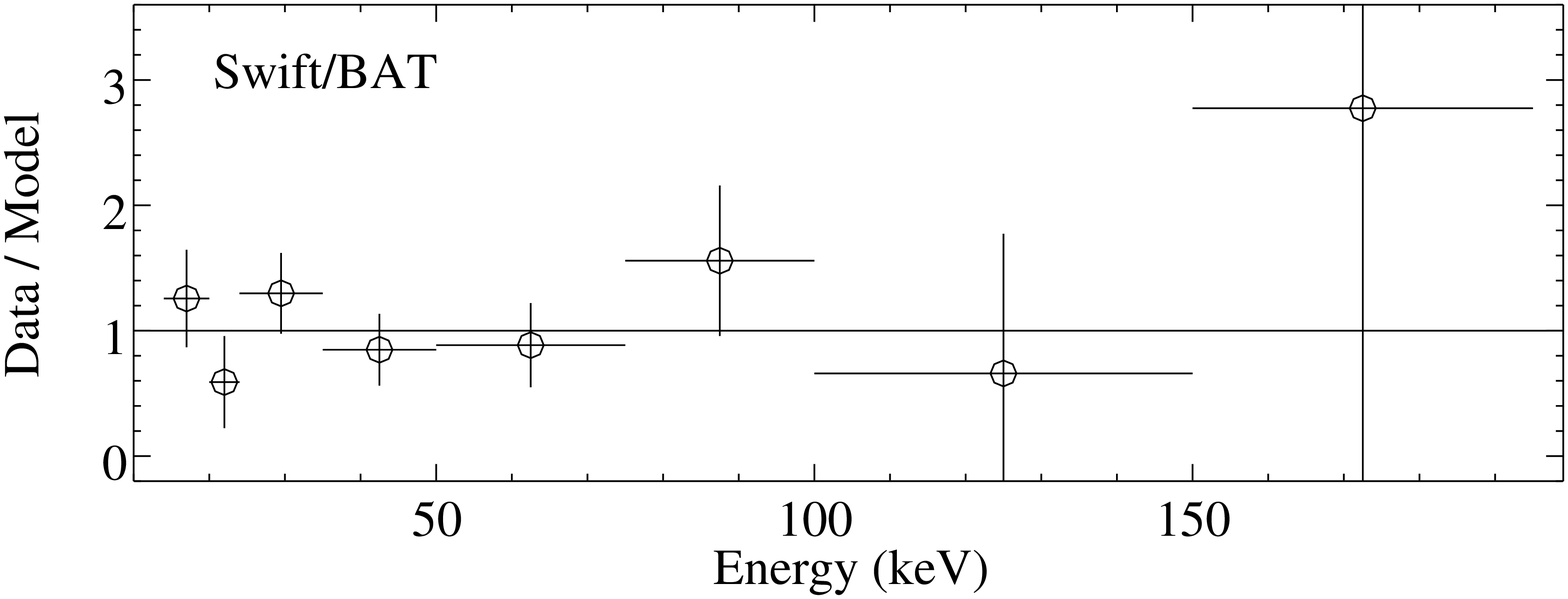}\\
\includegraphics[bb=80 -50 1055 282,scale=0.25]{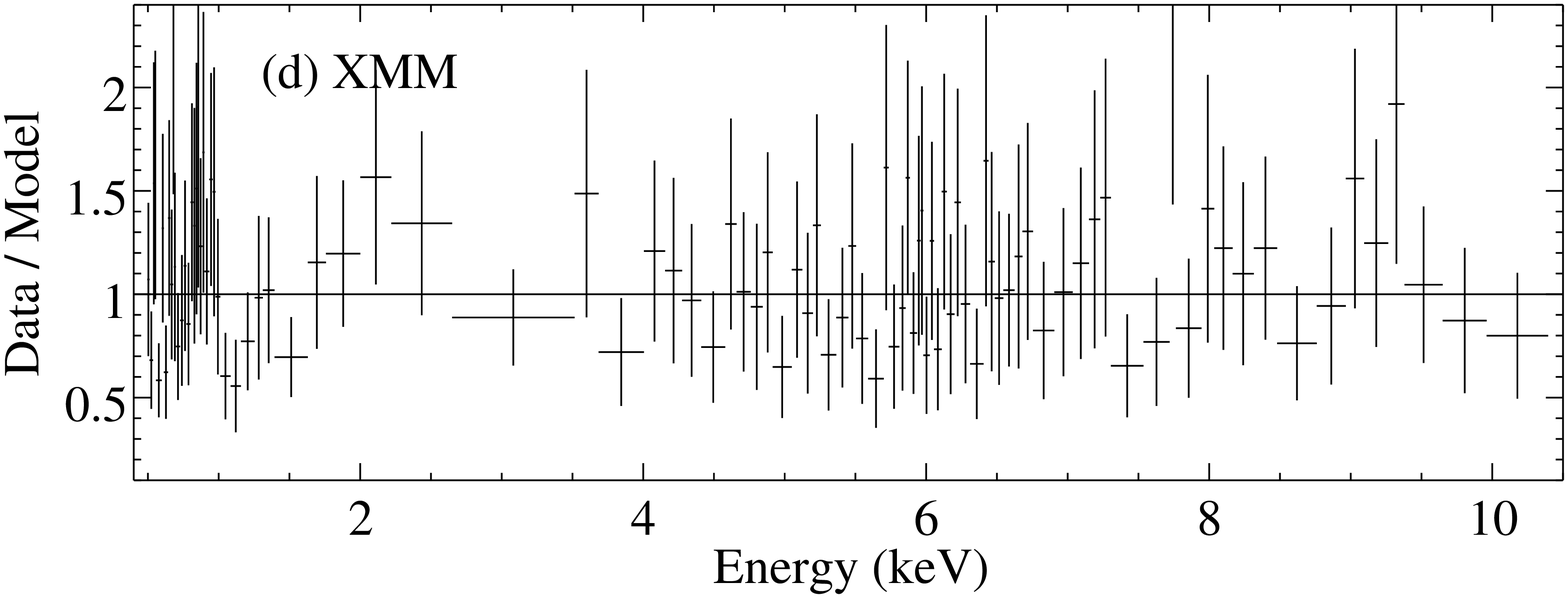}
\includegraphics[bb=80 -50 1055 282,scale=0.25]{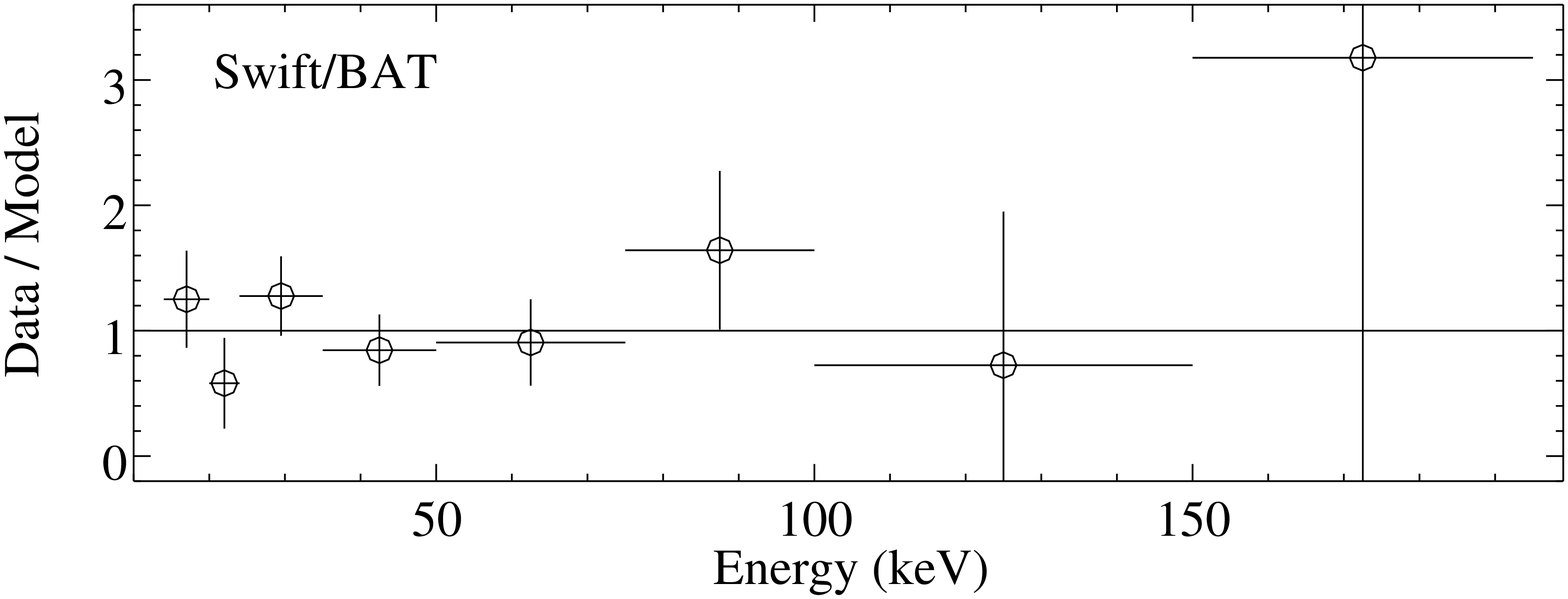}\\
\caption{Ratio of the {\it XMM-Newton} and {\it Swift}/BAT data to
(a) Model 3, fitting best the {\it Chandra} data;
(b) Model 3 with thawed temperature, $kT$, and normalization,
$N_{\rm TB}$, of the thermal bremsstrahlung component;
(c) Model 3a with thawed $N_{\rm TB}$, $kT$, and intrinsic absorption
column density, $N_{\rm H,\,z}$; left: {\it XMM-Newton}; right:
{\it Swift}/BAT;
(d) Model 3b, thawed $N_{\rm TB}$, $kT$, and normalization of the CPL,
$N_{\rm CPL}$; left: {\it XMM-Newton}; right: {\it Swift}/BAT.}
\label{fig.ratx}
\end{center}
\end{figure*}

\begin{figure*}
\begin{center}
\includegraphics[bb=68 10 580 495,clip,scale=0.36]{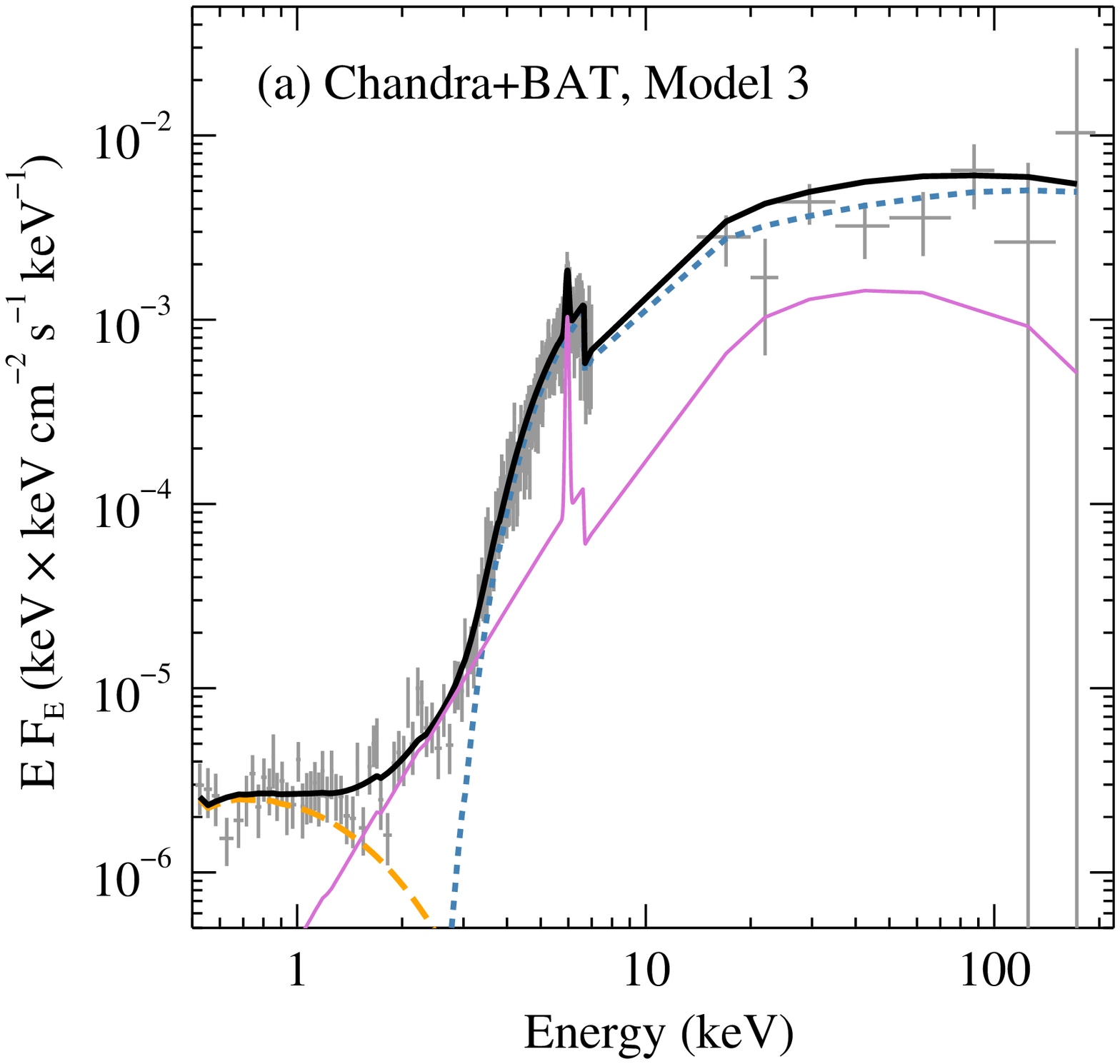}~
\includegraphics[bb=160 10 580 495,clip,scale=0.36]{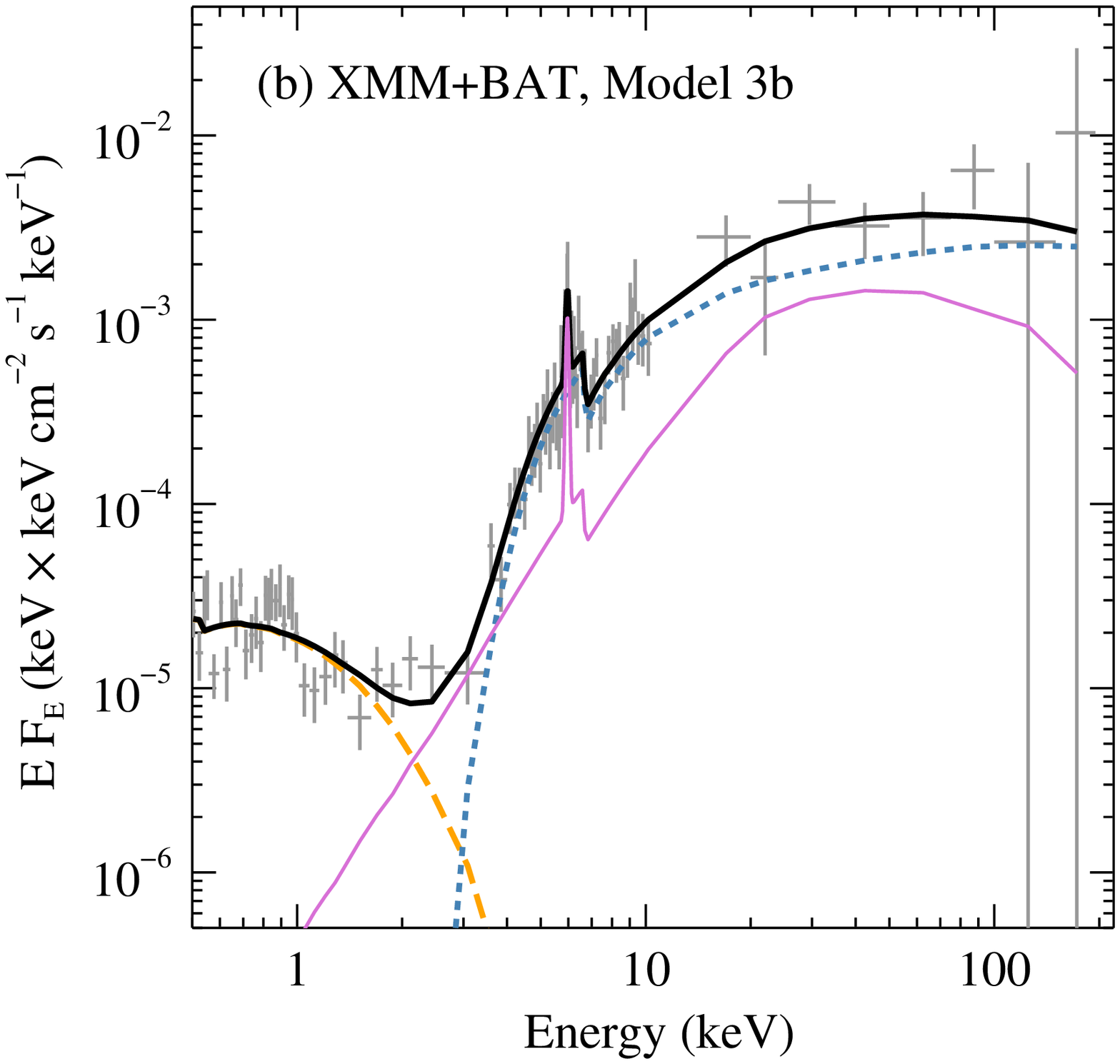}~
\includegraphics[bb=160 10 580 495,clip,scale=0.36]{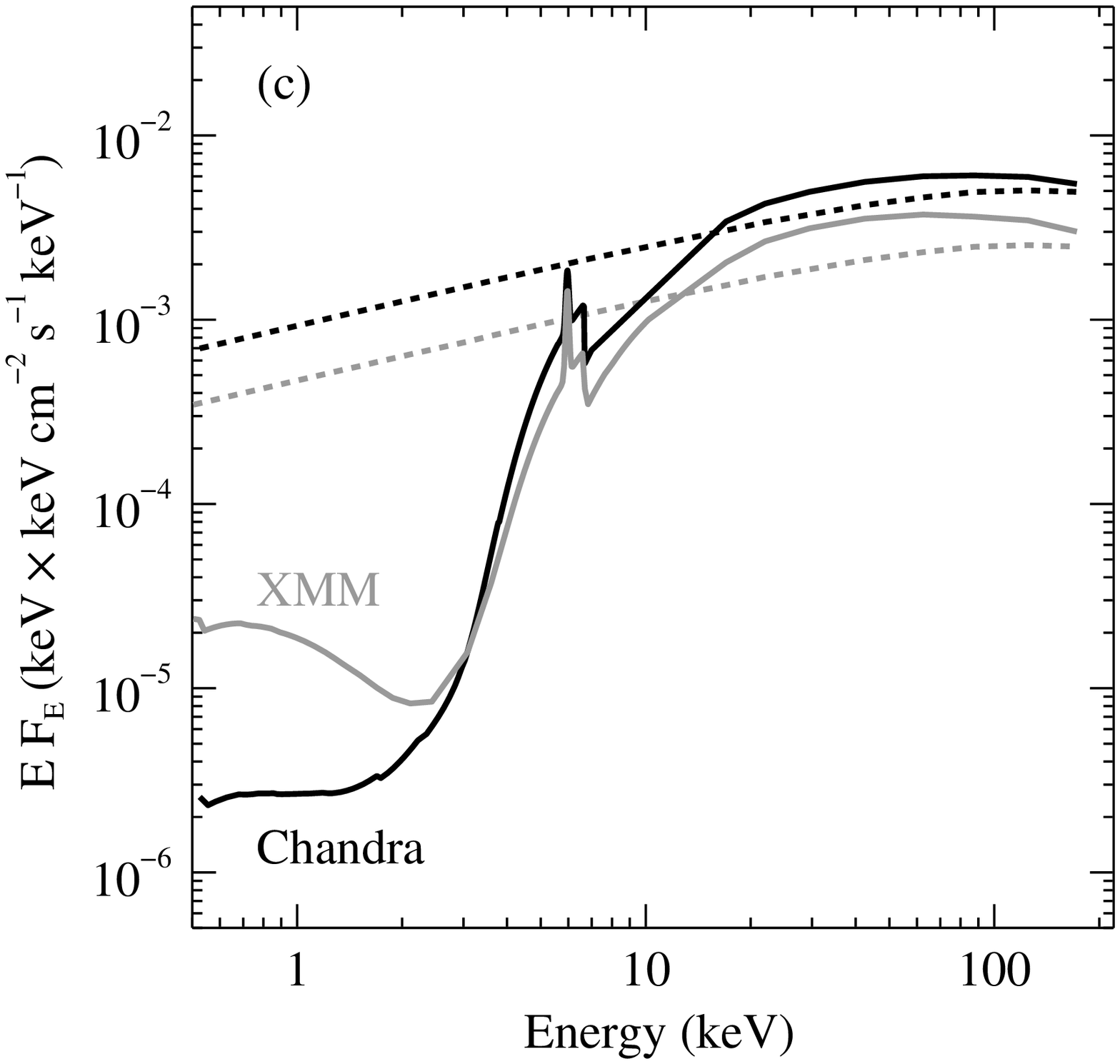}
\caption{(a) Unfolded spectrum ({\it Chandra} and {\it Swift}/BAT; crosses)
and best fitting model (Model 3; thick solid/black). The model components are as follows:
hard power-law emission (short-dash/blue) absorbed with an intrinsic column
density $N_{\rm H,\,z} \sim 5\times 10^{23}$\,cm$^{-2}$; reflected emission including
the $\sim 6.4$\,keV (rest frame) iron K$\alpha$ line (thin solid/magenta);
thermal diffuse emission (long-dash/orange); Galactic absorption. The {\it Chandra}
data were grouped for clarity of the plot.
(b) Same as (a) for {\it XMM-Newton} and {\it Swift}/BAT data (Model 3b).
(c) Solid -- best fitting {\it Chandra} (black) and {\it XMM-Newton} (gray) total models.
The enhanced soft diffuse emission in the {\it XMM-Newton} model is due to larger source extraction
region than in the case of {\it Chandra}. Short-dash -- unabsorbed power-law components with
variable normalization responsible for the intrinsic X-ray variability of the source. }
\label{fig.eeuf}
\end{center}
\end{figure*}

Figure~\ref{fig.ratx}a shows the ratio of the {\it XMM-Newton} data to
Model 3 with parameters fixed at the values fitting best the {\it Chandra} data.
A soft excess below 2\,keV is clearly visible, and is due to the diffuse emission of
gas surrounding the nucleus collected from a large extraction region required by the
{\it XMM-Newton} PSF.
To account for this difference we thawed the temperature and
normalization of the thermal bremsstrahlung component in Model 3
($kT$ and $N_{\rm TB}$, respectively). The fit in the soft band improved
significantly, although certain residuals indicative of emission/absorption
lines were still present. Since we are interested primarily in the continuum
AGN emission, we do not improve further on the soft ($<2$\,keV)
residuals and focus on modeling the spectrum at energies $>2$\,keV. The resulting
data-to-model ratios are plotted in
Figure~\ref{fig.ratx}b. The temperature $kT$ is lower than in the case
of the {\it Chandra} spectrum (although consistent within the
uncertainties) and close to the temperature of the thermal component in
the {\it Chandra} background model. Residuals below and above $\sim 4$\,keV
are still apparent indicating that the intrinsic spectral
properties of the nucleus might have changed between the {\it XMM-Newton} and
{\it Chandra} observations.  The shape of the residuals suggests that
either the intrinsic absorption column varied (model excess below 4\,keV)
or the normalization of the power-law varied (model deficit above 4\,keV).
We test both possibilities.

First we thaw the intrinsic absorption column (in addition to $kT$ and
$N_{\rm TB}$; Model 3a in Table~\ref{tab.mpar}) and repeat the fit. We
obtain an increased intrinsic column density, $N_{\rm H,\,z} \sim
8\times10^{23}$\,cm$^{-2}$, with respect to that in the {\it Chandra}
observation. The intrinsic EW is in agreement with that in {\it Chandra}
Model 3, as expected, while the observed EW increases to
EW$_{\rm o}=247 \pm 10$\,eV (Table~\ref{tab.mpar}, Figure~\ref{fig.ratx}c).
The fit resulted in ${\rm CSTAT}=228$ (191 d.o.f.).

Next, we test the hypothesis that the spectral variability is due to
the variable normalization of the power-law, $N_{\rm CPL}$ (Model 3b
in Table~\ref{tab.mpar}, Figure~\ref{fig.ratx}d),
while the intrinsic absorption and reflection parameters remain fixed
at the values obtained from the analysis of the {\it Chandra} data. To
correctly account for the constant absolute flux of the reflected
component, we release the condition that the normalization of the
reflected component is equal to the normalization of the
power-law. Instead, we fix the normalization of the reflected
component at the value found in the {\it Chandra} data.
We obtained ${\rm CSTAT}=209$ (191 d.o.f.). We find that $N_{\rm CPL}$ is
lower by a factor $\sim 2$ in the {\it XMM-Newton} data than in the {\it
Chandra} data. We note, however, that this parameter is subject to
large errors in the {\it Chandra} data. This spectral variability scenario results
in an increased equivalent width of the iron line (observed
EW$_{\rm o} = 280 \pm 13$\,eV, intrinsic EW$_{\rm i} = 122 \pm 6$\,eV) as
compared with the {\it Chandra} Model 3.
A combined variability scenario (i.e., both $N_{\rm H,\,z}$ and $N_{\rm CPL}$
allowed to vary; 190 d.o.f.) {resulted in parameter values and {\rm CSTAT}
consistent with those of Model 3b.}

Finally, we checked the consistency between the {\it XMM-Newton} and
{\it Swift}/BAT data. We found $C_{\rm BAT} = 0.68\pm0.10$ and
$C_{\rm BAT}=1.10\pm0.16$ in both cases, Model 3a and 3b, respectively.
The data-to-model ratios in the {\it Swift}/BAT energy range
are presented in Figures~\ref{fig.ratx}c--d.

\section{Discussion}

We studied the nuclear X-ray emission and X-ray spectral variability of a low-redshift
radio galaxy 4C+29.30 over a broad 0.5--200\,keV X-ray band using our deep $\sim 300$\,ksec
{\it Chandra} observation, $\sim 17$\,ksec of cleaned {\it XMM-Newton} data, and archival
58-months catalog {\it Swift}/BAT data. We tested several spectral models, and concluded that
Model 3 in Table~\ref{tab.mpar} provides the best fit to our {\it Chandra} data supplemented by
the {\it Swift}/BAT data. In this model the 2--200\,keV X-ray emission of the central AGN is
decomposed into a strongly intrinsically obscured hard cut-off power-law continuum ($\Gamma\simeq 1.56$),
and significantly less absorbed reprocessed emission including a neutral reflection hump and an
accompanying neutral narrow iron K$\alpha$ line. The soft 0.5--2\,keV X-ray band is dominated by
a thermal diffuse emission, and may contain a marginal contribution of a scattered power-law component.
The same model applied to the earlier {\it XMM-Newton} data led us to explore two X-ray variability
scenarios. In the first one, the variability is apparent and caused by the variations of the intrinsic column
density. In the other, the variability arises due to changes of the normalization of the 
power-law emission.

The hard X-ray spectrum, renewed nuclear radio activity (van Breugel et al. 1986;
Jamrozy et al. 2007; Liuzzo et al. 2009), and weak optical continuum dominated by the
host galaxy (van Breugel et al. 1986) suggest an analogy between the state of the nucleus in
4C+29.30 and a hard or hard-intermediate spectral state of Galactic black hole X-ray binaries
(GBHs; see e.g., Fender et al. 2004; Done et al. 2007). The spectral ages of large scale radio
structures in 4C+29.30 differ by $\sim 100$\,Myr. This timescale, when scaled down by the
ratio between the black hole mass in 4C+29.30 ($\sim 10^8$\,M$_{\odot}$, Section~\ref{sec:bhmass}
of the present paper; Paper I) and that of a typical GBH ($\sim 10$\,M$_{\odot}$), is
comparable with the time between consecutive outbursts of GBHs ranging from years to
decades (for examples of GBHs with at least two documented activity periods see e.g.,
Table~1 in Remillard \& McClintock 2006; Figure~2 in Done et al. 2007). Thus, it is possible
that we observed 4C+29.30 at an early stage of its renewed nuclear activity, which could
have started as recently as several decades ago, judging by the spectral age of the youngest
nuclear radio structures. Interestingly, the age of the youngest radio structure, $\sim 15$\,yr,
found by Liuzzo et al. (2009) is in agreement with the suggestion of Jamrozy et al. (2007) that
a powerful outburst took place in 4C+29.30 in the period between 1990 and 2005.
The X-ray loudness parameter (e.g., Zamorani et al. 1981; Avni \& Tananbaum 1986),
$\alpha_{\rm ox}\simeq 0.3838(\log F_{\rm \nu,\,o}-\log F_{\rm \nu,\,x})$, where $F_{\rm \nu,\,o}$
and $F_{\rm \nu,\,x}$ are rest-frame flux densities at 2500\,\AA\ and 2\,keV, respectively,
can be calculated for 4C+29.30 assuming the AGN contribution to the optical band as in van
Breugel et al. (1986), and 2\,keV flux corrected for absorption (Model 3). We find a low value,
$\alpha_{\rm ox} \simeq 1$, further supporting the idea of a hard or hard-intermediate
spectral state of the source (e.g., Sobolewska et al. 2009; 2011).

Below we discuss in detail the implications of our modeling on the nature of the studied AGN
and its surroundings.

\subsection{Origin of the X-ray spectral variability}

The three X-ray observations of 4C+29.30 available to date, spread over a period of several years,
allow us to address the question about X-ray variability of the source.
The early {\it Chandra} data (Apr 2001; Gambill et al. 2003) were described with a simple power-law
model ($\Gamma$ fixed at 1.69) modified by the Galactic absorption ($N_{\rm H}=4.06\times10^{20}$\,cm$^{-2}$,
slightly different than the value adopted in the present paper), and an intrinsic absorption
$N_{\rm H,\,z} = 4.8^{+1.9}_{-1.4}\times10^{23}$\,cm$^{-2}$ in excellent agreement 
with that found in our new {\it Chandra} data, $N_{\rm H,\,z} = (4.96\pm0.15)\times10^{23}$\,cm$^{-2}$.
The reported 2--10\,keV absorbed flux was $1.1 \times 10^{-12}$\,erg\,s$^{-1}$\,cm$^{-2}$.
Based on the Gambill et al. model parameters we evaluated that the unabsorbed 2--10\,keV flux of the
source in April 2001 was $F_{\rm C,\,1} \simeq 4.8 \times 10^{-12}$\,erg\,s$^{-1}$\,cm$^{-2}$,
comparable with that at the time of our {\it Chandra} observation,
$F_{\rm C,\,2} = 4.7^{+0.3}_{-1.2} \times 10^{-12}$\,erg\,s$^{-1}$\,cm$^{-2}$ (Feb 2010).
However, our deep observation revealed that the spectrum of the 4C+29.30 core is more complex due
to the soft component and reflection features. The model of Gambill et al. (2003) when applied to
our {\it Chandra} data results in a reduction of the 2--10\,keV flux, $F_{\rm C,\,2}$, by $\sim 6$\%.
It is thus possible that $F_{\rm C,\,1}$ in Gambill et al. (2003) was underestimated by a few percent.
Nevertheless, we conclude that the 2--10\,keV flux and the properties of the intrinsic absorber
were approximately unchanged between the two {\it Chandra} observations.

However, a discrepancy between our {\it Chandra} and {\it XMM-Newton} data (Apr 2008)
is apparent, as demonstrated in Figure~\ref{fig.ratx}. We proposed two scenarios for the X-ray
spectral variability of 4C+29.30. We started with the good quality deep {\it Chandra} observation to
establish a base-line spectral model, and we found that the {\it XMM-Newton} data above 2\,keV
can be explained by varying only one spectral parameter: the intrinsic absorbing column density,
$N_{\rm H,\,z}$, or the normalization of the power-law component, $N_{\rm CPL}$. In the former case
(variable $N_{\rm H,\,z}$), the {\it XMM-Newton} flux is in agreement with both {\it Chandra}
measurements, $F_{\rm N,\,a} = F_{\rm C,\,2} \simeq F_{\rm C,\,1}$, while the intrinsic column
density exceeds that measured with {\it Chandra} by $\sim 60$\%. In the latter case (variable
$N_{\rm CPL}$) we obtain $F_{\rm N,\,b} = (2.5 \pm 0.1) \times 10^{-12}$\,erg\,s$^{-1}$\,cm$^{-2}
\simeq 0.53 F_{\rm C,\,2}$. Cross-calibration
studies\footnote{\tt http://xmm.esac.esa.int/docs/documents/CAL-TN-0052.ps.gz}
show that in general the spectra measured by {\it XMM-Newton} (EPIC/PN) and {\it Chandra} (ACIS-S)
agree within 10\% in the 0.5--2\, keV band, and within 20\% above 2\,keV.
Thus, the detected {\it Chandra} and {\it XMM-Newton} intrinsic flux difference is too high
to be fully attributed to the cross-calibration issues, implying that in this scenario
the source may be variable on timescales of several years in the 2--10\,keV band.

Variations of the intrinsic column density seem to be a common phenomenon in AGN on
timescales as short as weeks (e.g., Risaliti et al. 2005, 2007, 2009, 2010).
On the other hand, a non-variable narrow 6.4\,keV iron line and associated constant
neutral reflection not responding to the variations of the
power-law component, interpreted as the reprocessing of the hard X-rays by a remote
molecular torus, also seem to be common spectral features in both Type 1 AGN (e.g., MCG-6-30-15,
Vaughan \& Fabian 2004; NGC 4051, Miller et al. 2010, Ponti et al. 2006; Mrk 841, Petrucci et al. 2007;
Ark 120, Nardini et al. 2011; a sample of 14 Seyfert 1 galaxies, Bhayani \& Nandra 2010)
and Type 2 AGN (e.g., a sample of 20 Seyfert 2 galaxies, Risaliti 2002; MCG-5-23-16,
Reeves et al. 2006; NGC 2992, Yaqoob et al. 2007, and references therein; NGC 4945, Itoh et al. 2008,
Marinucci et al. 2012).

In our modeling, both variability scenarios resulted in satisfactory fits
(Table~\ref{tab.mpar}). However, a visual inspection of the fit residuals in
the {\it XMM-Newton} and {\it Swift}/BAT energy ranges (Figure~\ref{fig.ratx})
seems to favor the scenario with variable power-law normalization, $N_{\rm CPL}$ (Model 3b)
over the scenario with variable intrinsic column density, $N_{\rm H,z}$ (Model 3a).
We confirmed that model with variable $N_{\rm CPL}$ is more likely to be correct
than model with variable $N_{\rm H,z}$ by calculating the respective p-values
(Protassov et al. 2002) based on the series of 1000 simulated data sets (using Sherpa's
{\tt fake\_pha} function) for two test cases: (a) variable $N_{\rm H,z}$ (null model) vs. combined
variability scenario (both $N_{\rm H,z}$ and $N_{\rm CPL}$ allowed to vary (alternative model),
and (b) variable $N_{\rm H,z}$ (null model) vs. combined variability scenario (alternative model).
In case (a) we found p-${\rm value} \ll 1 \times 10^{-3}$, which implied that model
with variable $N_{\rm H,z}$ is unlikely to be true. Conversely, in case (b) the high
p-${\rm value} = 0.32$ suggests that the model with variable $N_{\rm CPL}$ is likely
to be true.

In Figure~\ref{fig.eeuf}c we show a comparison between the best fitting {\it Chandra}
and {\it XMM-Newton} total models, and the unabsorbed power-law model components in the
case of the favored scenario with variable $N_{\rm CPL}$.

\subsection{Iron line and reflection hump}

The first short {\it Chandra} observation of 4C+29.30 contained too few counts to detect
the iron K$\alpha$ line or put a limit on its EW (Gambill et al. 2003). Our {\it XMM-Newton}
and {\it Chandra} observations are thus the first detections of the K$\alpha$ iron line
in this object. The line was narrow ($\Delta E\sim 0.06$\,keV) and centered at the rest
frame energy close to 6.4\,keV. These line properties indicate that it originated in a neutral
matter far away from the black hole, i.e. outer parts of an accretion disk or a dusty torus.

The {\it intrinsic} equivalent width of the iron line, EW$_{\rm i}$, measured in our {\it XMM-Newton}
and {\it Chandra} data is in the $\sim 70$--120\,eV range, depending on the adopted variability
scenario (Table~\ref{tab.mpar}). An iron line originating in a lamp-post model due to reprocessing
of hard power-law X-rays with $\Gamma=1.9$ by an accretion disk observed face-on has an expected
EW of $\sim 140$\,eV (e.g., George \& Fabian 1991). However, this EW is sensitive to a number
of parameters such as the photon index, viewing angle, iron abundance, and optical depth for
electron scattering. Nandra et al. (2007) give the photon index and inclination dependent
formulas for the EW based on the fits to the Monte Carlo simulations of George \& Fabian (1991).
For $\Gamma\sim1.56$ resulting from our fits and an assumed inclination of 45$^\circ$, we obtain
the predicted EW of $\sim$170\,eV. This value should be considered as an upper limit due to the
unknown disk/X-ray source geometry (the reflection amplitude in our data is $|\Omega/2\pi|\sim0.3$ which
may suggest, e.g., a truncated disk geometry or a patchy corona, as opposed to $\Omega/2\pi=1$ in the
lamp-post model). It is thus possible that at least some of the photons forming an iron line with an
(intrinsic) EW of 70--120\,eV, as measured in this paper, originated in an accretion disk. This would require that
the intrinsic absorber and the reflector were two separate media.

Ghisellini et al. (1994) presented a scenario in which the iron line is formed
as a result of the reprocessing of a
power-law continuum by a molecular torus (see also Krolik et al. 1994; Levenson et al. 2002).
The torus scenario for the origin of the iron line provides a compelling possibility
of associating the intrinsic absorber with the reflector. The EW of the iron line for tori with
$N_{\rm H} \sim (5-8)\times 10^{23}$\,cm$^{-2}$ and a half-opening angle of 30$^{\circ}$--45$^{\circ}$
is predicted in the $\sim 80$--300\,eV range, with the lower (higher) values corresponding to
the face-on (edge-on) view. Ghisellini et al. (1994) assumed an illuminating power-law with
$\Gamma=1.9$. Levenson et al. (2002) estimated that the allowed range of EWs increases by up
to 5\% if the spectrum hardens with $\Gamma=1.9$ changing to $\Gamma=1.7$ (for intermediate viewing
angles $i<65^\circ$). The spectrum measured in our data is even harder, characterized with $\Gamma\sim1.56$,
which implies even larger allowed EWs. Comparing these predictions with the {\it observed} iron
line EWs in 4C+29.30, we find that both {\it Chandra} and {\it XMM-Newton} measurements,
EW$_{\rm o} \sim 170$--280\,eV, lie within the range predicted by the torus origin of the iron line.

In Model 3, we used the {\tt pexriv+zgauss} model of X-ray reprocessing.
We are aware of certain issues associated with this approach, which include the iron line added to
a continuum in an arbitrary way, and the assumption that the hard X-rays are reflected by an
accretion disk. We argue instead that the reflector could be associated with a molecular torus.
Models of hard X-ray reprocessing by a toroidal reflector are available  (Ghisellini et al. 1994;
Krolik et al. 1994; Levenson et al. 2002; Murphy \& Yaqoob 2009). Of these, the tabulated model
{\tt mytorus} of Murphy \& Yaqoob (2009) accounts self-consistently for all components reprocessed by an
optically thick torus -- the zero-order absorbed component, the scattered component, and associated fluorescent
Fe K$\alpha$, Fe K$\beta$ and Ni K$\alpha$ emission lines. However, the model in its present version considers
a torus with a fixed half-opening angle of $\theta_{\rm MYT}=60^{\circ}$. This model tested on our {\it Chandra}
data requires almost an edge on view, $i=85^{\circ}$, in contradiction with the viewing angle constrains
based on the jet-to-counterjet flux ratio. The data of 4C+29.30 require that the zero-order component is
strongly absorbed, indicating that the torus half-opening angle $\theta < i < 60^{\circ}$.
With viewing angle fixed at $i=45^{\circ}$, an additional intrinsic absorber affecting only
the zero-order component of the {\tt mytorus} model is needed in order to obtain a good fit,
which breaks the consistency of the model. We conclude that the available {\tt mytorus} model does not
provide at this time enough flexibility to account for the case appropriate for our source, in which
$\theta < i < \theta_{\rm MYT}=60^{\circ}$.

\subsection{4C+29.30 -- a `hidden' AGN}

Recently, a new class of AGN has been identified in the {\it Suzaku} and {\it XMM-Newton}
follow-ups of the {\it Swift}/BAT AGN with $N_{\rm H} > 10^{23.5}$\,cm$^{-2}$
(e.g., Ueda et al. 2007; Winter et al. 2008, 2009a,b; Eguchi et al. 2009, 2011; Noguchi et al.
2009, 2010). The characteristic feature of these AGN is a lower contribution of the scattered
power-law emission to the soft X-ray band, $f_{\rm sc} < 3\%$,
as compared to that of typical Seyfert 2 galaxies, $f_{\rm sc} = 3$--10\% (e.g., Turner et al. 1997;
Guainazzi et al. 2005; Winter et al. 2009a; Noguchi et al. 2009, 2010). Winter et al. (2009a)
found that 24\% of their entire {\it Swift}/BAT
selected sample (153 sources) belong to this new class of AGN based on the $f_{\rm sc} < 3\%$
and low soft-to-hard X-ray observed flux ratio, $F_{\rm 0.5-2\,keV}/F_{\rm 2-10\,keV} < 0.04$.
Ueda et al. (2007) proposed that the reduced $f_{\rm sc}$ is due to a geometrically thick torus
covering a significant fraction of the solid angle. They predicted that in this new class of AGN,
the [OIII] lines relative to hard X-ray luminosity would be weaker than in the typical Seyfert 2
galaxies because of lower flux arriving to the NLR in the presence of a geometrically thick torus.
A name `hidden' or `buried' AGN was proposed for this new class, missed in previous optically
and soft X-ray selected samples.

Noguchi et al. (2009, 2010) identified hidden AGN candidates in the 2XMM Catalogue using
X-ray hardness ratios, HR3 and HR4 (for definition see the 2XMM User Guide to the Catalogue).
These two hardness ratios calculated for the {\it XMM-Newton} data of 4C+29.30, HR3=0.09 and
HR4=0.83, comply with the Noguchi et al. selection criteria. In our analysis, we
found that the scattered component was not formally required to model the data and that the
scattering fraction was consistent with zero (Model 6),
$f_{\rm sc} < 0.25 \times 10^{-2}$. This value is quite low but in agreement with
$f_{\rm sc}$ found previously in several other Seyfert 2s (Ueda et al. 2007; Eguchi et al. 2009;
Noguchi et al. 2009; Winter et al. 2009a,b), and implies a torus with half opening angle
$\theta\lesssim 20^\circ$ (Eguchi et al. 2009 for $f_{\rm sc} < 0.5\%$). The observed
soft-to-hard flux ratio in 4C+29.30 is $F_{\rm 0.5-2\,keV}/F_{\rm 2-10\,keV} \sim 0.005$. Thus,
the {\it XMM-Newton} colors, $f_{\rm sc}$, and soft-to-hard flux ratio indicate that 4C+29.30 is
a hidden AGN candidate.

Noguchi et al. (2010) showed that the scattering fraction $f_{\rm sc}$ correlates with the ratio of
the reddening corrected [OIII] $\lambda$5007\AA\ line luminosity to the 2--10\,keV intrinsic luminosity
$\ell=L_{\rm [OIII]}/L_{\rm 2-10\,keV}$. Thus, to further exploit the intriguing hidden AGN scenario
in 4C+29.30, we investigated the properties of its [OIII] emission line. We used the observed [OIII]
flux reported by van Breugel et al. (1986). We corrected it for the extinction using their observed
${\rm H}_\alpha/{\rm H}_\beta$ luminosity ratio of 5.1 and the relation of Bassani et al. (1999) with the
assumed Galactic gas to dust ratio, i.e. $({\rm H}_\alpha/{\rm H}_\beta)_{\rm 0}=3.1$. We obtained $\ell\sim 0.1$ in 4C+29.30,
which means that our source can be placed somewhat above the Noguchi et al. correlation, however within the
scatter produced by the sources used to derive the correlation.

Base on a sample of six AGN, Eguchi et al. (2009) speculated that hidden AGN with
$f_{\rm sc} < 0.5\%$ have high reflection amplitudes $\Omega/2\pi\gtrsim0.8$, while those
with $0.5\% < f_{\rm sc} < 3\%$ have $\Omega/2\pi\lesssim0.8$. 4C+20.30 with $|\Omega/2\pi|=0.32^{+0.09}_{-0.05}$
and $f_{\rm sc} < 0.25\%$ fits in the gap between the two groups in the $\Omega/2\pi$ vs.
$f_{\rm sc}$ diagram (similarly to two other sources in Eguchi et al. 2011), indicating that the
distribution of $f_{\rm sc}$ and $\Omega/2\pi$ may be continuous rather than split into two distinct groups.

The soft thermal emission identified in our {\it Chandra} image contributes majority of the 0.5--2\,keV flux.
However, this component may be easily overlooked in the data with lower exposure or spatial resolution.
We investigated a possible effect of neglecting this component in the total model (Model 7 in Table~\ref{tab.mname}).
We found that the most striking difference between Model 3 and Model 7 is the softening of the photon index in the
case without the thermal component, $\Gamma_{\rm M7} = 2.03\pm0.16$. This follows from the adjustment of the slope
of the scatterred unabsorbed component (linked in the model to that of the CPL) in order to account for
the soft X-ray emission but not to overestimate the hard X-rays at the same time.

We conclude that more detailed studies of the soft emission in the hidden AGN complemented
by better quality hard X-ray data (e.g. NuSTAR) are of particular importance to recognizing their
X-ray properties and geometry of the obscuring torus.

\subsection{Black hole mass}
\label{sec:bhmass}

Noguchi et al. (2010) showed that the scattering fraction, $f_{\rm sc}$, anti-correlates
with the Eddington luminosity ratio, $L_{\rm bol}/L_{\rm E}$, where
$L_{\rm bol} \simeq 30\times L^{\rm CPL}_{\rm 2-10\,keV}$ (applying a bolometric correction
factor of 30 typical for luminous AGN; e.g., Vasudevan \& Fabian 2007). For 4C+29.30 the limit
we found on the scattering fraction translates into $L_{\rm bol}/L_{\rm E} \gtrsim 0.1$, which
gives $M_{\rm BH} \lesssim 1.2\times10^8$\,M$_{\odot}$.

Another black hole mass estimate can be done using the empirical relation between
$M_{\rm BH}$ and stellar velocity dispersion, $\sigma_{\star}$ (Tremaine et al. 2002).
Based on the optical spectrum taken from the Sloan Digital Sky Survey Data Release 7
(Abazajian et al. 2009) the stellar velocity dispersion of 4C+29.30 is
$\sigma_{\star} = 207.6$ km s$^{-1}$ (using the STARLIGHT synthesis code of Cid
Fernandes et al. 2005), resulting in the black hole mass $M_{\rm BH}=1.6\times10^8$\,M$_{\odot}$.

These two new black hole mass estimates are compatible with the discussion presented
in Paper I where $M_{\rm BH} \gtrsim 10^8$\,M$_{\odot}$ was found based
on an argument involving energetics of the system. Thus, it is possible that the mass of
the black hole in 4C+29.30 is close to $10^8$\,M$_{\odot}$.

\subsection{Multiwavelength perspective of 4C+29.30}

\subsubsection{Infra-red}

X-ray radiation absorbed and reprocessed by an obscuring torus would
eventually emerge in the infrared band. Models of the infrared
AGN SEDs under the assumption of a clumpy structure of a molecular torus
(Nenkova et al. 2002; H{\"o}nig et al. 2006; Nenkova et al. 2008a,b) are
potentially able to constrain the torus geometry in sources observed in IR
with a high spatial resolution (e.g., Gandhi et al. 2009; Alonso-Herrero et al. 2012),
with the near-IR 1--5\,$\mu$m fluxes being most sensitive to the inclination effects
(Nenkova et al. 2008b). 4C+29.30 was observed in infrared by IRAS, 2MASS, and WISE, of which
the two latter surveys sampled the desired near-IR band. However, in 4C+29.30
the 2MASS measurements are dominated by the host galaxy\footnote{We estimated
the host galaxy contribution to the $J$, $H$ and $K_s$ bands using the flux
measurement at 5400\AA\ being a sum of $\lesssim 80$\% and $\gtrsim 20$\% galaxy
and AGN emission, respectively (van Breugel et al. 1986); the $E(B-V) = {\rm 0.5}$\,mag
reddening possibly due to a dust lane (van Breugel et al. 1986); and the 10\,Gyr
elliptical galaxy template of Buzzoni (2005).}, while the WISE observations
with a spatial resolution 6$\arcsec$--12$\arcsec$ correspond to our 10$\arcsec$
{\it Chandra} background region rather then the core region. Nevertheless,
we use the W2 and W3 filters at rest-frame wavelengths $\lambda_{\rm W2} \sim 4.2\,\mu$m
and $\lambda_{\rm W3} \sim 11.3\,\mu$m, to constrain the viewing angle of our
source. We calculated respective $f = \nu F_{\nu}/F_{\rm bol}$ ratios (where $F_{\rm bol}$
corresponds to $L_{\rm bol} \sim 10^{45}$\,erg\,s$^{-1}$) and compared them with the
4.5\,$\mu$m and 12\,$\mu$m models of Nenkova et al. (2008b; Figure~10; see also Erratum in Nenkova et al. 2010).
We obtained $f_{\rm 4.3\,\mu m} \simeq 0.06$ and $f_{\rm 11.3\,\mu m} \simeq 0.08$, which gave $i
\simeq 70^\circ$ and no constraint (our $f_{\rm 11.3\,\mu m}$ was too low to intercept the
12\,$\mu$m model of Nenkova et al. 2008b), in contradiction with the $i<60^\circ$ condition followed
from the jet-to-counterjet flux ratio. The $f_{\rm 4.3\,\mu m}$ and $f_{\rm 11.3\,\mu m}$ ratios
are a factor 2--6 too low to comply with the $i<60^\circ$ constraint. There are several reasons for
this discrepancy to consider. (i) The bolometric luminosity of 4C+29.30 might have been underestimated
by a factor of a few due to the uncertainty of the applied bolometric correction. (ii) The geometry of
the obscuring torus may differ from that considered in Nenkova et al. (2008b), e.g., the case of
a geometrically thick torus with half-opening angle $\lesssim 20^\circ$ evoked for hidden AGN by
Eguchi et al. (2009) would imply a torus width parameter $\sigma \gtrsim 70^\circ$ (see Figure~1 in
Nenkova et al. 2008b for the illustration of $\sigma$) as opposed to
$\sigma=45^\circ$ assumed to compute the models. (iii) Finally, the $\sim 12\,\mu$m WISE flux might
be partially affected by the 4C+29.30 ISM environment (see e.g., Goulding et al. 2012 for arguments
in favor of host galaxy origin of the rest-frame 9.7\,$\mu$m Si-absorption feature in Compton-thick
Seyfert 2 galaxies), e.g., a dust lane covering the central part of the nucleus, clearly visible in the
archival HST image in Figure~\ref{fig:opt}.

We verified the third scenario using the correlation reported in Gandhi et al. (2009) for a number
of Seyfert 2 galaxies between their 2--10 keV luminosities and 12\,$\mu$m luminosities. The IR
luminosities in Gandhi et al. (2009) follow from sub-arcsecond resolution observations,
and thus most likely they are representative of the true IR nuclear emission. In 4C+29.30 the
$\sim 12\,\mu$m luminosity is $L_{12\,\mu{\rm m}} \sim8.3\times10^{43}$\,erg\,s$^{-1}$. This is
in agreement with the expected $L_{12\,\mu{\rm m}} \sim8.8^{+2.2}_{-1.8}\times10^{43}$\,erg\,s$^{-1}$
(based on our unabsorbed 2--10\,keV {\it Chandra} flux; Model 3), leaving scenarios (i) and/or (ii)
as plausible causes of the discrepancy between the Nenkova et al. (2008b) models and 4C+29.30 WISE flux
measurements.

\subsubsection{Radio}

4C+29.30 is an intermittent radio galaxy with evidence for several jet ejection events with the spectral time
ranging from $\gtrsim200$\,Myr (the relic structure), through $\lesssim100$\,Myr and  $\lesssim33$\,Myr, to
$\sim 15$\,yr and $\sim 70$\,yr (Jamrozy et al. 2007; Liuzzo et al. 2009).
However, the radio activity of 4C+29.30 seems rather mild in comparison with the powerful radio sources.
The total jet power of 4C+29.30, $L_j \sim 10^{42}$\,erg\,s$^{-1}$, constitutes only
a small fraction of its Eddington luminosity, $L_j \sim 10^{-4} L_{\rm E}$, while in the most powerful blazars
$L_j$ reaches (0.1--$1)L_{\rm E}$ (e.g., Ghisellini et al. 2009). Low-power FR I radio galaxies
are characterized by much lower values of $L_j/L_{\rm E}$ (e.g., $\sim 10^{-3}$ in the case of M87),
but at the same time they accrete at low rates ($L_{\rm bol}/L_{\rm E} \ll 0.1$) probably in
a radiatively inefficient mode. To the contrary, we estimated that the Eddington luminosity ratio of our
low-power radio galaxy is quite substantial, $L_{\rm bol} \gtrsim 0.1 L_{\rm E}$.

The standard radio loudness parameter is typically defined as the ratio of 5\,GHz radio to B-band (4400\AA)
optical luminosity spectral densities, $R \equiv L_{\nu,\,{\rm R}} / L_{\nu,\,{\rm B}}$, with $R>10$
characterizing radio-loud AGN (as proposed for quasars by Kellermann et al. 1989). In the case of 4C+29.30,
one can calculate $90 \lesssim R \lesssim 430$ based on the available archival data (5 GHz total flux given
in Liuzzo et al. 2009 and 4400\AA\ flux extrapolated from the allowed range of the 5400\AA\ nuclear flux,
assuming $\alpha_{\rm opt}=1.2$; see the discussion in van Breugel et al. 1986).
However, this value should be taken with caution mainly because of uncertainties associated with the
attenuation of the optical nuclear flux due to the obscured nature of the source, and host galaxy contribution
known to be particularly relevant in the case of low-power Seyfert galaxies (Ho \& Peng 2001; Sikora et al. 2007).

We compared the radio properties of 4C+29.30 with those of 52 hidden AGN candidates compiled from
the literature based on Ueda et al. (2007), Eguchi et al. (2009, 2011), Noguchi et al. (2009, 2010),
and Winter et al. (2009a,b). In approximately one-third of them (17 of 52) the scattered fraction was
reported to be $\lesssim 0.5$\% in at least one publication, that is at a level comparable with the upper
limit estimated in 4C+29.30, implying a geometrically thick torus with an opening angle $\lesssim$20$^\circ$
(Eguchi et al. 2009). Combination of the available archival NVSS (Condon et al. 1998) and
NED data allowed us to calculate the standard radio loudness parameter in the case of 34 sources,
including 14 of 17 sources with $f_{\rm sc} < 0.5$\%. We found that approximately one-third of these sources
(11 of 34) could be considered radio-loud according to the condition $R>10$. Among them we distinguished five
radio-intermediate sources with $R \sim 10$--20,
and six sources with a substantially higher radio loudness, $R \sim (2$--27$)\times10^3$.
We caution, however, that similarly to the case of 4C+29.30, the calculated values of the radio loudness are only
approximate due to the obscured nature of the hidden AGN. Nonetheless, it seems that the hidden nuclei are associated
predominantly with the radio-quiet AGN.

Finally, we stress that among the 34 sources with measured radio loudness, $R$, we found only three ($\sim 9$\%) with
$f_{\rm sc} < 0.5$\% and $R > 10$ (Mrk 348, 3C 33, and NGC 612). Hence, 4C+29.30 appears quite unusual in
that it shows properties characteristic to both radio-loud AGN and hidden AGN with extremely geometrically thick
tori. Based on the present data it cannot be determined whether a small fraction of radio-loud AGN among
the hidden AGN population is due to physical constraints or an observational bias.

\section{Conclusions}

We confirmed that the X-ray emission of the AGN in 4C+29.30, a radio source hosted by an elliptical galaxy,
resembles Seyfert 2 type of activity. We estimated that the mass of the central black hole is of the order
of 10$^8$\,M$_{\odot}$. We decomposed the broad band 0.5--200\,keV X-ray continuum ({\it Chandra}/{\it XMM-Newton}
supplemented by {\it Swift}/BAT observations) into
(i) a hard ($\Gamma = 1.5$--1.6) cut-off power-law component attenuated due to a strong intrinsic
X-ray absorption with column density $N_{\rm H} \sim 5 \times 10^{23}$\,cm$^{-2}$;
(ii) a reflection hump ($|\Omega/2\pi|\sim0.3$) and narrow iron K$\alpha$ line detected
for the first time in this source originating due to the reprocessing of the AGN power-law emission
by a distant neutral matter;
(iii) a soft thermal emission of a diffuse gas clearly detected in our {\it Chandra} image
dominating the 0.5--2\,keV band;
(iv) a marginal contribution of a scattered power-law emission to the soft X-ray band,
$f_{\rm sc} < 0.25$\%.
We hypothesized that the distant reflector and the obscuring matter can be both associated with
a molecular torus. Based on $f_{\rm sc}$, the ratio of the observed soft-to-hard X-ray fluxes, and the
{\it XMM-Newton} colors we proposed that 4C+29.30 belongs to the hidden/buried AGN, a new class that is
emerging among the {\it Swift}/BAT hard X-ray selected AGN. We demonstrated that 4C+29.30, a radio-loud
source with an intermittent jet activity and $f_{\rm sc}$ below 0.5\%, shows properties characteristic
to only $\sim 9$\% of the presently identified hidden/buried AGN candidates with estimated radio
loudness parameter.\\

\acknowledgments
We thank the anonymous referee for careful reading of our manuscript
and comments that led to its improvement. \L .S. is grateful for the
support from Polish MNiSW through the grant N-N203-380336. M.J. was
supported by Polish MNiSW funds for scientific
research in years 2009--2012 under the contract No. 3812/B/H03/2009/36.
Work at NRL (C.~C.~C.) is sponsored by NASA DPR S-15633-Y. This research has
made use of data obtained with the Chandra X-ray Observatory, and
{\it Chandra} X-ray Center (CXC) in the application packages CIAO, ChIPS,
and Sherpa.  This research is funded in part by NASA contract
NAS8-39073. Partial support for this work was provided by the
{\it Chandra} grants, GO0-11133X and GO1-12145X, and XMM-Newton
grant NNX08AX35G.



\begin{sidewaystable*}
\begin{center}
\caption{XSPEC/Sherpa terminology of the spectral models.}
\label{tab.mname}
\begin{tabular}{ll}
\hline\hline
Name & Description\\
\tableline
Model 1$^a$ & {\tt phabs *} ({\tt zbremss} + {\tt zphabs*}CPL)\\
Model 2 & {\tt phabs *} ({\tt zbremss} + {\tt zphabs*}CPL + {\tt zgauss})\\
Model 3 & {\tt phabs *} ({\tt zbremss} + {\tt zphabs*}CPL + {\tt zgauss} + {\tt pexriv})\\
Model 4 & {\tt phabs *} ({\tt zbremss} + {\tt zphabs*}CPL + {\tt zgauss} + {\tt zphabs*pexriv})\\
Model 5$^b$ & {\tt phabs *} ({\tt zbremss} + {\tt zpcfabs*}CPL + {\tt zgauss} + {\tt zphabs*pexriv})\\
Model 6$^c$ & {\tt phabs *} ({\tt zbremss} + {\tt zphabs*}CPL + {\tt constant*}CPL + {\tt zgauss} + {\tt zphabs*pexriv})\\
Model 7$^c$ & {\tt phabs *} ({\tt zphabs*}CPL + {\tt constant*}CPL + {\tt zgauss} + {\tt zphabs*pexriv})\\
\hline
\end{tabular}\\
\end{center}
NOTE -- Models 4--6 converged to Model 3. In {\tt Sherpa} the individual model names are proceeded by {\tt xs}.\\
$^a {\rm CPL} \equiv {\tt zhighect*zpowerlw}$.\\
$^b {\tt zpcfabs} \equiv f_{\rm pc} {\tt *zphabs} + (1-f_{\rm pc})$.\\
$^c {\tt constant} \equiv f_{\rm sc}$.\\

\begin{center}
\caption{Results of spectral fits to the {\it Chandra}, {\it XMM-Newton} and {\it Swift}/BAT data.}
\label{tab.mpar}
\begin{tabular}{lccccccccccccc}
\hline\hline
& \multicolumn{2}{c}{Thermal} & Intrinsic & \multicolumn{2}{c}{Cut-off power-law$^b$} & \multicolumn{4}{c}{Fe K${\alpha}$ line$^c$} & Neutral & Norm.$^d$ & \multicolumn{2}{c}{CSTAT}\\ 
& \multicolumn{2}{c}{bremsstrahlung$^a$} & absorption & \multicolumn{2}{c}{(CPL)} & \multicolumn{4}{c}{} & reflection &  &  & \\
\hline
 & $kT$ & $N_{\rm TB}$ & $N_{\rm H,\,z}$ & $\Gamma$ & $N_{\rm CPL}$ & $E$ & $\Delta E$ & EW$_{\rm o}$ & EW$_{\rm i}$ & $|\Omega/2\pi|$ & $C_{\rm BAT}$ & Value & Prob. (d.o.f.)\\
 & keV &  & $10^{23}$\,cm$^{-2}$ &  &  & keV & keV & eV & eV &  &  &  & \\
\hline\hline
 & \multicolumn{13}{c}{{\it Chandra}}\\
\hline
Bkg & $0.63\pm0.04$ & $3.01^{+0.30}_{-0.27}$ & ... & $-2.28\pm0.22$ & $0.0035^{+0.0016}_{-0.0011}$ & ... & ... & ... & ... & ... & ... & 578 & 1 (442)\\
Model 1  & $6.4^{+6.5}_{-2.1}$ & $0.46^{+0.04}_{-0.02}$ & $4.08\pm0.29$
 & $0.84^{+0.11}_{-0.27}$ & $27^{+18}_{-11}$ & ... & ... & ... & ... & ... & $0.20\pm0.03$ & 1129 & $\sim 10^{-7}$ (887)\\ 
Model 2 & $5.2^{+3.0}_{-1.5}$ & $0.46\pm0.02$ & $3.91^{+0.16}_{-0.09}$
 & $0.97^{+0.07}_{-0.03}$ & $31^{+21}_{-12}$ & $6.36\pm0.02$ & $0.07\pm0.03$ & $192^{+54}_{-37}$ & $91^{+27}_{-18}$ & ... & $0.30\pm0.05$ & 1045 & $\sim 10^{-4}$ (884)\\
Model 3 & $0.71^{+0.15}_{-0.12}$ & $0.83^{+0.21}_{-0.16}$ & $4.96\pm0.14$ &
$1.56^{+0.15}_{-0.04}$ & $103^{+39}_{-50}$ & $6.36\pm0.01$ & $0.06\pm0.03$ & $173^{+62}_{-35}$ & $71^{+27}_{-15}$ & $0.32^{+0.05}_{-0.13}$ & $0.68\pm0.10$ & 1016 & 0.002 (883)\\
\hline
 & \multicolumn{13}{c}{{\it XMM-Newton}}\\
\tableline
Bkg & $1.62^{+2.24}_{-0.81}$ & $1.34^{+0.45}_{-0.20}$ & ... & $-0.37^{+0.35}_{-0.48}$ & $0.13^{+0.13}_{-0.08}$ & ... & ... & ... & ... & ... & ... & 1219 & 1 (1903)\\
Model 3a & $0.60\pm0.07$ & $8.5^{+1.4}_{-1.1}$ & $7.88\pm0.26$ & $1.56^*$ & $103^*$ & $6.36^*$ & $0.06^*$ & $247 \pm 10$ & 64 & $0.32^*$ & $0.68\pm0.10$ & 228 & 0.03 (191)\\
Model 3b & $0.60\pm0.07$ & $8.7^{+1.4}_{-1.1}$ & $4.96^*$ & $1.56^*$ & $52.0\pm{2.7}$ & $6.36^*$ & $0.06^*$ & $280 \pm 13$ & $122 \pm 6$ & $0.32^*$ & $1.10\pm0.16$ & 209 & 0.30 (191)\\
\hline
\end{tabular}\\
\end{center}
$^a$ Normalization of the thermal bremsstrahlung, $N_{\rm TB}$, in units
$10^{-5} \times [3.02\times10^{-15}/(4\pi D^2 (1+z)^2)]\times\int n_e n_i {\rm dV}$,
where $n_e$ is the electron density in cm$^{-3}$, $n_i$ is the ion density in cm$^{-3}$, $D$ is the angular size distance in cm,
and $z$ is the redshift. \\
$^b$ Normalization of the cut-off power-law, $N_{\rm CPL}$, in units 10$^{-5}$\,photons\,keV$^{-1}$\,cm$^{-2}$ s$^{-1}$ at 1\,keV.\\
$^c$ The observed (intrinsic) equivalent width EW$_{\rm o}$ (EW$_{\rm i}$) calculated relative to the intrinsically absorbed (unabsorbed) continuum.\\
$^d$ Normalization constant between the {\it Chandra} ({\it XMM-Newton}) and {\it Swift}/BAT models.\\
$^*$ Parameter fixed at the Model 3 value.\\

\end{sidewaystable*}

\end{document}